\documentclass{aa}
\usepackage{graphicx}
\usepackage{txfonts}
\usepackage[colorlinks=true,citecolor=blue]{hyperref}
\usepackage{xspace}

\begin{document}

   \title{VVVX near-IR photometry for 99 low-mass stars in the \emph{Gaia} EDR3 Catalog of Nearby Stars\thanks{Tables 1 - 5 are only available in electronic form at the CDS via anonymous ftp to \url{cdsarc.u-strasbg.fr} (130.79.128.5) or via \url{http://cdsweb.u-strasbg.fr/cgi-bin/qcat?J/A+A/}}}

   \subtitle{}

   \author{A. Mej\'ias
          \inst{1,}
          \and
          D. Minniti
          \inst{1,2}
          \and
          J. Alonso-Garc\'ia
          \inst{3,4}
          \and
          J. C. Beam\'in
          \inst{5}
          \and
          R. K. Saito
          \inst{6}
          \and
          E. Solano
          \inst{7,8}
          }

          \institute{Departamento de Ciencias F\'isicas, Universidad Andr\'es Bello, Av. Fern\'andez Concha 700, Santiago, Chile\\
              \email{andreac.mejiasa@gmail.com}
         \and
             Vatican Observatory, V00120 Vatican City State, Italy
             %\email{}
             %\thanks{}
          \and
             Centro de Astronom\'ia (CITEVA), Universidad de Antofagasta, Av. Angamos 601, Antofagasta, Chile
          \and
             Millennium Institute of Astrophysics, Nuncio Monse\~nor Sotero Sanz 100, Of. 104, Providencia, Santiago, Chile
          \and
             N\'ucleo de Astroqu\'imica y Astrof\'isica, Instituto de Ciencias Qu\'imicas Aplicadas, Facultad de Ingenier\'ia, Universidad Aut\'onoma de Chile, Av. Pedro de Valdivia 425, Santiago 7500912, Chile
           \and
              Departamento de F\'isica, Universidad Federal de Santa Catarina, Trindade 88040-900, Florian\'opolis, SC, Brazil
            \and
               Departamento de Astrof\'isica, Centro de Astrobiolog\'ia (CSIC-INTA), ESAC Campus, Camino Bajo del Castillo s/n, E-28692 Villanueva de la Ca\~nada, Madrid, Spain
            \and
                Spanish Virtual Observatory, E-28692 Villanueva de Ca\~nada, Madrid, Spain
                }

   \date{Received 12 July, 2021 ; accepted 13 December, 2021}

% \abstract{}{}{}{}{} 
% 5 {} token are mandatory
 
  \abstract
   {Red dwarf stars, which represent 75\% of stars in the Milky Way, can be studied in great detail in the solar neighborhood, where the sample is more complete.}
   {We intend to better characterize red-dwarf candidates selected from the \emph{Gaia} Catalog of Nearby Stars (GCNS) using optical and near-infrared multi-filter photometry from the Vista Variables in the V\'ia L\'actea eXtended (VVVX) Survey, the DECam Plane Survey (DECaPS), the Panoramic Survey Telescope and Rapid Response System (Pan-STARRS), and the Wide-field Infrared Survey Explorer (WISE).}
   {We performed a cross-matching procedure among the positions of a color-selected sample of M dwarfs in the VVVX Survey and the \emph{Gaia} Early Data Release 3 sub-catalog of nearby stars. We explored their stellar parameters and spectral types using the Virtual Observatory SED Analyzer (VOSA). Radii were also obtained from the computed luminosities and T$_{eff}$ using the Stefan-Boltzmann equation. Masses and ages were computed for some of the objects using evolutionary tracks and isochrones. Additional mass estimations were obtained with the $M_{Ks} - M_{*}$ relation. We then validated our results for the stellar parameters of two of our objects with spectra obtained with the TripleSpec instrument at the SOAR telescope, as well as those of our total amount of stars through a direct comparison with an independent sample from the literature. We revised the objects in our sample and compared their proper motion vectors with other sources within 30'' to identify possible companions and probed their renormalized unit weight error (RUWE) values to identify unresolved companions.}
   {We present a catalog of physical parameters for 99 low-mass objects with distances from 43.2 to 111.3 pc. Effective temperatures range from 2500 to 3400 K, with the majority of stars in the sample compatible with the status of M4 dwarfs. We obtained a good agreement between the stellar parameters computed with VOSA and the estimations from observed spectra, also when comparing with an independent sample from the literature. The distribution of masses obtained with VOSA is concentrated toward the very low-mass regime. Eight objects present values of RUWE $\geq$ 1.4 and seven are consistent with being part of a binary system.}
   {}


\titlerunning{99 VVVX low-mass stars in the \emph{Gaia} Catalogue of Nearby Stars}
\authorrunning{Mej\'ias et. al}
\maketitle
%
%________________________________________________________________

\section{Introduction}
\label{int}

Low-mass stars, also known as red dwarfs or M dwarfs, are the lowest mass stars that are capable of fusing hydrogen in their cores (0.075 $<$ M$_{\odot}$ $<$ 0.6), with spectral types between M0 to M9. They comprise around 75\% of the objects in the Milky Way, and $\sim$ 50\% of these stars may harbor terrestrial exoplanets, with one in seven M dwarfs hosting Earth-sized planets within the habitable zone \citep{2013ApJ...767L...8K, 2015ApJ...807...45D}. Their low masses and faint luminosities maintain the habitable zone, defined as the region where water can exist in a liquid state on a nearby exoplanet's surface \citep{1964hpfm.book.....D, 1979Icar...37..351H, 1993Icar..101..108K}. In addition, when compared to Sun-like stars or earlier types, the smaller masses and radii of red dwarfs favor the detection and characterization of smaller and less massive exoplanets via transit and radial velocity methods, since the light curves and radial velocity signals are more significant for Earth-sized planets around red dwarfs compared with the signals produced by the same planet around a larger star \citep{2021AJ....161..247F, 2011PASP..123..709M}.  

Therefore, planets that are orbiting within the habitable zone are more easily discovered and characterized, making M dwarfs primary targets for exoplanet search missions. It is then particularly interesting to identify and characterize M dwarfs that are close to the Sun, since their stellar parameters will be better constrained, allowing us to better determine the properties of an exoplanet orbiting them. 

%%LITERATURA PREVIA
Over the last twenty years, several efforts have been made to study samples of red dwarfs in the vicinity of the Sun. In the work of \citet{2004AJ....128..463R}, the New Luyten Two Tenths \citep[NLTT;][]{1980nltt.bookQ....L} proper motion catalog was cross-matched with the Two Micron All-Sky Survey (2MASS) Second Incremental release to produce a catalog of 1913 stars, of which 1910 have astrometric, photometric, or spectroscopic distances (or a combination of the three), and 815 of those stars are likely to be within 20 pc of the Sun. This resulting catalog from the NLTT follow-up observations is composed of M dwarfs, spanning spectral types between M2 and M7.
Large-scale surveys have also allowed for the identification of very low-mass stars and sub-stellar objects as brown dwarfs. In 2008, \citet{2008MNRAS.383..831P} reported nearby M and L dwarfs within 30 pc at low Galactic latitudes ($|b|<15^{o}$) over 4800 square degrees in the Deep Near-Infrared Southern Sky Survey (DENIS) database. Another example is given the work of \citet{2013PASP..125..809T}, where the authors presented 42 low-mass stars and brown dwarf candidates, including 15 M dwarfs with Wide-Field Infrared Survey Explorer (WISE) satellite photometry. 

Other catalogs such as the All-Sky Catalog of Bright M dwarfs from \citet{2011AJ....142..138L} provide information on 8889 stars selected from the SUPERBLINK survey of stars with proper motions of $\mu<40$ mas yr$^{-1}$. The catalog provides parallax measurements (when available) and photometric distance estimates for all stars, placing most of the sources within 60 pc of the Sun. Spectral types range from K7 to M4, with later spectral types only within 20 pc.

Low-mass M dwarfs are fully convective and represent an important case study focused on magnetic activity. In the work of \citet{2021AJ....161...63W}, the authors explored this mass regime and presented a volume-complete, all-sky list with 512 M dwarfs within 15 pc with masses between $0.1 \leq M/M_{\odot} \leq 0.3$. 

%GAIA
With the \emph{Gaia} mission, the European Space Agency (ESA) seeks to make the largest, most precise map of our Galaxy down to G $\sim$ 20 mags, becoming more precise and accurate with each data release. This allows for the determination of distances to different objects in the Milky Way, as well as studies of the structure of our Galaxy and our solar neighborhood in a five-dimensional (5D) space when considering the proper motions in right ascension and declination \citep{2016A&A...595A...1G}. 

In \emph{Gaia} Second Data Release (DR2) there are 3050 ultra-cool dwarfs ($\geq$ M7), 647 L, and 16 T dwarfs that have been spectroscopically confirmed. In the work of \citet{2018A&A...619L...8R}, these objects were used to identify new ultra-cool and brown dwarf candidates in the \emph{Gaia} DR2 by comparing their locus in the Hertzsprung-Russel diagram. This yielded $\sim$ 13 000 dwarfs with subtypes later than M7 and 631 new L candidates, demonstrating that \emph{Gaia} offers a great opportunity to probe these objects, which can then be further characterized with new photometric and spectroscopic observations. A characterization of the end of the main sequence and ultra-cool dwarfs in the first and second Gaia data release was performed by \citet{2017MNRAS.469..401S, 2019MNRAS.485.4423S}.

Compared to the previous two Data Releases \citep{2016A&A...595A...2G, 2018A&A...616A...1G}, the new \emph{Gaia} Early Third Data Release (EDR3), reduces the random and systematic errors in parallaxes by another 30\% and increases the number of objects from $\sim$ 1.7 billion objects to $\sim$ 1.8 billion objects \citep{2021A&A...649A...1G}. The \emph{Gaia} Catalog of Nearby Stars \citep[GCNS;][]{2021A&A...649A...6G} contains a census of $331,312$ objects in the solar neighborhood, which is defined  as a sphere of 100 pc centered on the Sun. This is a volume-complete sample for all objects earlier than M8 at the nominal G = 20.7 magnitude limit of \emph{Gaia}.

This new data release also triggered the compilation of all the stars and brown dwarfs within 10 pc of the Sun \citep{2021A&A...650A.201R}. Based on this, the authors estimated that around 61\% of the objects are M stars and more than half of the M stars are within the range from M3.0V to M5.0V. In total, the catalog contains 540 objects spanning stars, brown dwarfs, and exoplanets belonging to 339 systems.

%VVVX Survey
With the VISTA Variables in the V\'ia Lactea eXtended Survey \citep[VVVX;][]{2018ASSP...51...63M}, we can provide new and deep near-infrared photometry for these nearby objects at low Galactic latitudes and deeper in the Galactic plane. Previous studies have also probed the central regions of our Galaxy in the aim to identify low-mass objects. In particular, in the inner regions of the Galaxy, \citet{2013A&A...557L...8B} identified the first brown dwarf of the VISTA Variables in the Via Lactea Survey (VVV). It was characterized as an L5$\pm$1, an unusually blue dwarf. On the other hand, \citet{2015MNRAS.454.4476S}, identified a brown dwarf companion to the A3V star $\beta$-Circini in a proper motion and parallax catalog of the VVV survey. This is an unusually wide system (6656 AU), and the brown dwarf is estimated to have an age of 370-500 Myr and a mass of 0.056$\pm$0.007 M$_{\odot}$. Known high proper-motion stars were also studied with this survey. \citet{2013A&A...560A..21I} looked for new nearby object companions of these stars, exploring the multi-epoch images of the 2MASS and VVV surveys to search around 167 known high proper motion stars and identified seven new co-moving companions and discarded a pair of stars that were shown not to be co-moving after all. On the other hand, \citet{2018RMxAC..50...55B} used the VVV data to measure precise astrometry for eighteen known high proper motion sources, and detected five systems that are most likely very low-mass stars belonging to the Galactic halo. 

In this work, we additionally provide deep optical photometry in the \emph{grizY} bands from the DECAM Plane Survey \citep[DECaPS;][]{2018ApJS..234...39S} and the Panoramic Survey Telescope And Rapid Response System Survey \citep[Pan-STARRS;][]{2016arXiv161205560C}. Due to the importance of precise characterizations of the sources to accurately determine the properties of any possibly orbiting exoplanets, we computed effective temperatures and spectral types and probed their astrometric parameters to identify objects that could be in binary systems.

The paper is organized as follows. In Section \ref{data}, we describe the different surveys where we obtained the multi-wavelength photometry to select low-mass stars in the Galactic plane. In Section \ref{selection}, we provide a summary of the selection method for low-mass stars in the VVVX footprint. We then present the methods that we use to determine their stellar parameters and spectral types from photometry only, along with a further characterization from the spectroscopic data for two of our objects in Section \ref{param}. The results of our computations and their validations can be found in Section \ref{results}, followed by a discussion in Section \ref{discussion}, and the main conclusions of our work are presented in Section \ref{conclusions}.

%__________________________________________________________________

\section{Data}
\label{data}

In this work, we explored a region of the southern Galactic disk with the VVVX survey \citep{2018ASSP...51...63M}, an extension of the VVV survey that maps the midplane and bulge of the Milky Way in three near-infrared photometric bands (\emph{JHKs}). The VVVX survey fills the gap between the VVV and VISTA Hemisphere Survey (VHS) and expands the coverage of the sky from 562 deg$^2$ in VVV to $\sim 1700$~deg$^2$. It also extends the VVV time-baseline enabling proper motion measurements of $\lesssim 0.3$ masyr$^{-1}$ in the optically obscured regions where \emph{Gaia} is limited by extinction \citep{2018MNRAS.474.1826S}. Our region of study corresponds to an area of 120$^{\circ}$ x 9$^{\circ}$, consisting of two stripes of 83 x 2 tiles each, where one tile corresponds to a field of view of 1.501 deg$^2$.

A VVVX photometric catalog is being produced (Alonso-Garc\'ia et al., in prep.) following the point spread function (PSF) fitting techniques used to create the VVV PSF photometric catalog, as described in \citet{2018A&A...619A...4A}. We used a preliminary version of this new catalog to identify the sources and their near-infrared photometry in our region of interest.

Astrometric information was obtained from the Second Data Release (DR2) of the \emph{Gaia} mission and was updated with the new Early Third Data Release (EDR3; see Table 1).

Optical photometry data were obtained from the DECaPS Survey \citep{2018ApJS..234...39S} for those regions between $240^{\circ} > l > 5^{\circ}$, and from Pan-STARRS1 \citep{2016arXiv161205560C} for $240^{\circ} > l > 230^{\circ}$. The DECaPS Survey is a five-band optical and near-infrared survey of the southern Galactic plane with the Dark Energy Camera (DECam) at Cerro Tololo. The total footprint of this survey covers $\sim$ 1000 $deg^{2}$. On the other hand, the Pan-STARRS1 \citep[PS1;][]{2016arXiv161205560C} is the first part of Pan-STARRS that has already been completed and forms the basis for both Data Releases 1 and 2 (DR1 and DR2). The PS1 survey used the 1.8 Pan-STARRS Telescope \#1 in Hawaii and its 1.4 Gigapixel camera to image the sky in five broadband filters \emph{grizY}.

Finally, we added the WISE magnitudes reported in the \emph{Gaia} Catalog of Nearby Stars when available to obtain infrared photometry at longer wavelengths than the ones of the VVVX survey. WISE mapped the entire sky at the beginning of 2010 and continued through early 2011 in four mid-infrared bandpasses centered at wavelengths of 3.4, 4.6, 12, and 22 $\mu$m namely W1 to W4, respectively \citep{2010AJ....140.1868W}. WISE’s two shortest wavebands, W1 and W2, are the most valuable in searching for cold stars and brown dwarfs because these objects emit a substantial amount of their flux at these wavelengths, which are the ones used in this work. These values and the near-infrared photometry data of our objects can be found in Tables 2 and 3.

\section{Selection of low-mass objects}
\label{selection}

The 99 objects included in this work are a sub-sample of red dwarfs candidates that were color-selected in the VVVX survey disk region. The procedure for the selection of these stars is briefly described below, and a full description of our selection method will be presented in Mejías et al. (in prep).

First, we performed a cross-match between the positions reported for the VVVX sources with DECaPS or Pan-STARRS depending on the region of the VVVX footprint, and then with the \emph{Gaia} DR2 data to obtain astrometric information as parallaxes and proper motions. The specific steps done for the selection of red dwarf candidates are as follows: 1) We applied a quality criteria to the photometry, selecting only sources with $\sigma_{K_{s}} < 0.1$. 2) From this high-quality data, we selected all sources with a parallax value larger than 1.0 mas. 3) To select our low-mass stars candidates, we applied near-IR and optical color cuts, and 4) we used the reduced proper motion value to discriminate possible red giant contaminants.

To determine the color cuts, we explored a spectroscopically confirmed sample of M dwarfs from the work of \citet{2011AJ....141...97W}. We determined the mean ($J-H$), ($H-K_{s}$) and ($J-K_{s}$) colors and defined our color cuts to be within 1-$\sigma$ from the mean values (Eqs. \ref{nir-cuts1}-\ref{nir-cuts3}):

\begin{align}
0.414 < J &- H < 0.695, \label{nir-cuts1} \\
0.058 < H &- Ks < 0.504, \label{nir-cuts2} \\
0.621 < J &- Ks < 1.051. \label{nir-cuts3}
\end{align}

On the other hand, we based our optical color cuts in those reported in the work of \citet{2009AJ....138..633K} (Eqs. \ref{vis-cuts1}-\ref{vis-cuts3}):

\begin{align}
&i - z \geq 0.3, \label{vis-cuts1} \\
&r - i \geq 0.6, \label{vis-cuts2} \\
&r - i \geq 1.35*[(i-z)-0.319] + 0.6. \label{vis-cuts3}
\end{align}
   
In a color-selected sample of M dwarfs, red giant stars are the main contaminants. We carried out a reduced proper motion selection to identify possible giant stars that could be present in the sample. We calculated the J magnitude reduced proper motion for the objects in the sample with the following definition:

\begin{equation}
   \label{reduced_pm}
   \begin{split}
        H_{J} = J + 5log_{10} \mu
   \end{split}              
\end{equation}

We used the same criteria adopted in \citet{2014A&A...571A..36R}, where dwarf stars are all the objects with a reduced proper motion greater than the result given by the Eq. \ref{reduced_pm_sel}:

\begin{equation}
   \label{reduced_pm_sel}
   \begin{split}
        H_{J}^{dwarf} > H_{J}^{*} = 68.5(J-Ks) - 50.7
   \end{split}              
.\end{equation}

The total sample contains $N \sim 3.7$ $\times$ $10^{5}$ objects, updated with the outcome of the EDR3. Distances were also updated to the EDR3 distances computed from a Bayesian approach by \citet{2021AJ....161..147B}.

\begin{figure*}[h]
   \centering
    %\hspace*{-0.5cm}
    \includegraphics[width=\textwidth]{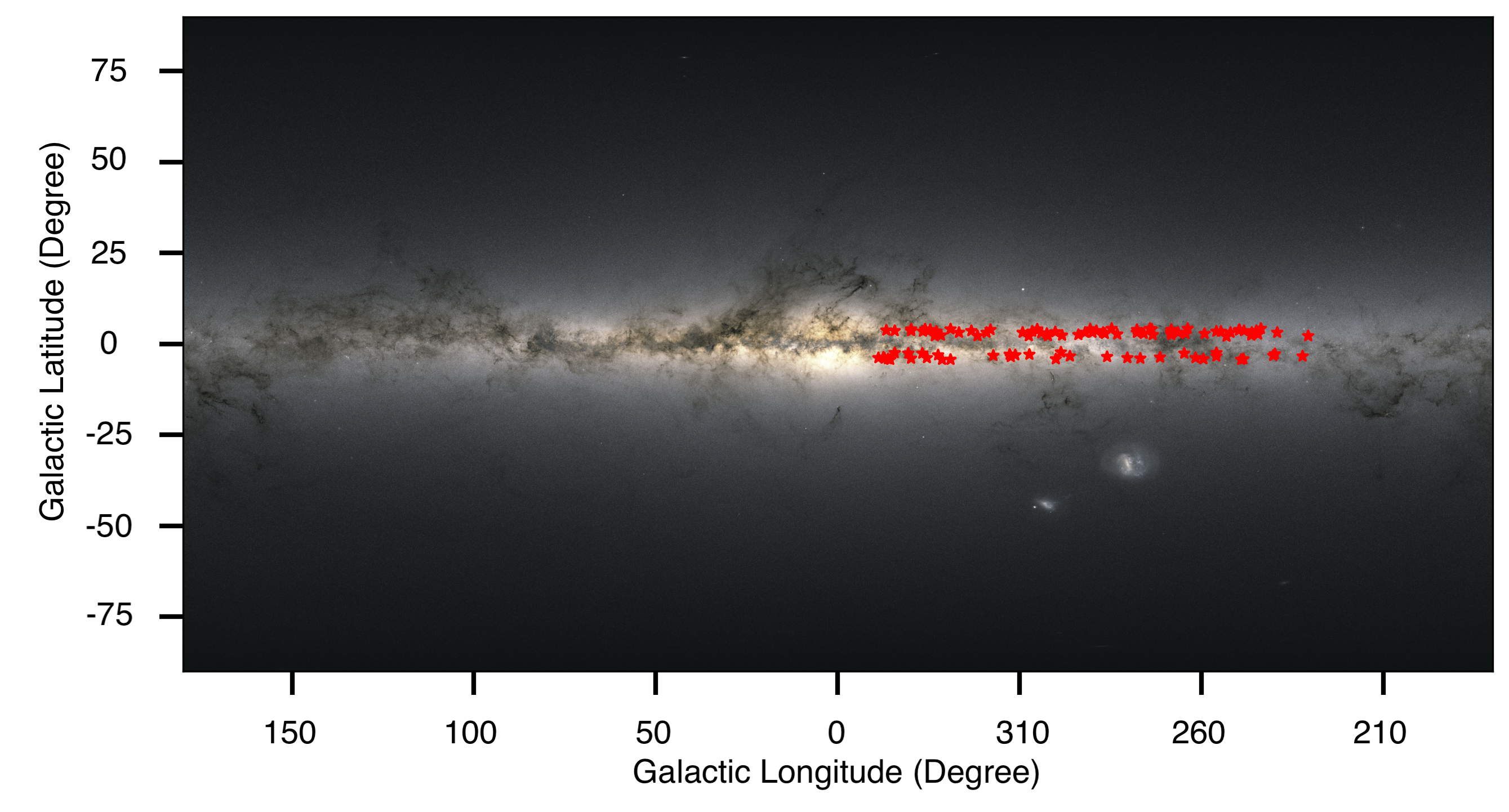}
    \caption{The 99 objects in the VVVX Survey footprint included in the \emph{Gaia} Catalog of Nearby Stars overplotted (in red) over a full-sky image of 6500 x 3250 px generated by ESA \emph{Gaia} EDR3. The sources of this work are distributed all over the region included in the work of Mejías et al. (in prep.) from $230^{\circ} \leq l \leq 350^{\circ}$ and $\pm4,5^{\circ} \leq b \leq \pm 2^{\circ}$.}
    \label{sky_gcns}
\end{figure*}

The recent Early Data Release of the Gaia Mission includes a smaller catalog of $331,312$ objects that are within 100 pc of the Sun \citep{2021A&A...649A...6G}. We performed a positional cross-matching with a 1'' search radius between our sample and the \emph{Gaia} Catalog of Nearby Stars with and found 99 objects, shown in Fig. \ref{sky_gcns}.

For the 99 low-mass objects of this work, distance values typically do not change between DR2 and EDR3 when they are closer than $\sim$ 80 pc. For larger distances, EDR3 values indicate that objects are closer by 1.4\% and more precise (31\% smaller errors). On the other side, proper motions are 0.2\% higher with 55\% smaller errors, corresponding to an important improvement of the astrometric parameters for the objects in our sample. 

\section{Stellar parameters}
\label{param}

\subsection{Effective temperatures, log(g), radii, masses, and ages}
\label{seccion41}
Determinations of astrophysical parameters is fundamental to understanding the properties of stars. We used the Virtual Observatory SED Analyzer \citep[VOSA;][]{2008A&A...492..277B} to estimate effective temperature and surface gravity, assuming [Fe/H] = 0.0 for our objects, which is a consistent value for stars in the solar neighborhood \citep{2001MNRAS.325..931R, 2001MNRAS.325.1365H, 2011A&A...530A.138C}, together with estimations of mass and radius. These estimations are done by fitting theoretical models from a library that is available in the VOSA database. 

The procedure is summarized as follows: we first provided the photometry of our objects in the optical bandpasses of the DECaPS and Panstarrs Surveys (\emph{grizY}), along with the near-infrared bands of the VVVX Survey (\emph{JHK$_{s}$}). We additionally added WISE photometry when available.

\begin{figure}[h]
\centering
%\hspace*{-0.3cm}
\includegraphics[width=\linewidth]{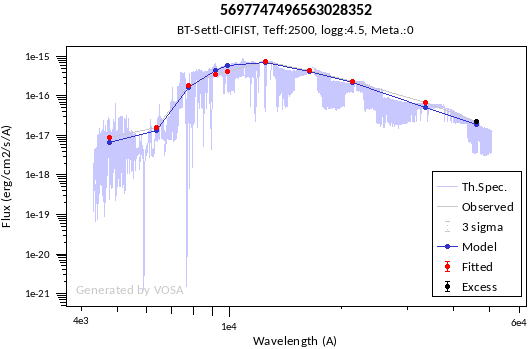}
\caption{Spectral energy distribution of the object \object{EDR3 5697747496563028352}. The best-fit model yields an effective temperature of 2500 K, with a value of log $g$ of 4.5. Red points show the photometry of the object in our optical, near-infrared, and infrared bands (\emph{grizY JHK$_{s}$ W1W2}). In blue, the best-fit model is shown, together with the theoretical spectrum that is plotted behind.} 
\label{sed}
\end{figure}

Then we used the VOSA to construct the spectral energy distribution of each object, which allows the user to select the collection of theoretical models to be used in the fitting. The fit is made through a chi-square test between the model and the SED of the object (see Fig. \ref{sed}).

In this work, we also made use of the BT-Settl CIFIST grid of models \citep{2011ASPC..448...91A}, where a solar metallicity value is adopted, together with the following range in effective temperature and surface gravity: 

\begin{center}
1200 K $\leq$ T$_{eff}$ $\leq$ 4500 K, \\
4.5 $\leq$ log $g$ $\leq$ 5.5.  
\end{center}

The radii of our objects are calculated using the Stefan-Boltzmann equation, making use of the luminosities and effective temperatures previously computed by VOSA. We also estimated the masses and ages for our objects with the isochrones and evolutionary tracks available in the VO. For each object, VOSA uses the theoretical isochrones and evolutionary tracks appropriate for the model that best fits the observed photometry. In our case, since we used the BT-Settl CIFIST models to fit the spectral energy distributions of the objects of this work, VOSA computes the ages and masses with the \citet{2015A&A...577A..42B} isochrones and evolutionary tracks. To do so, the values of effective temperature and luminosities obtained from the previous chi-square fit are used again, but this time as starting points for interpolating the theoretical isochrones and evolutionary tracks. 

\subsection{Spectral types}
\label{seccion42}
To compute the spectral types of these objects, we fitted empirical stellar spectra from the spectral library of \citet{2017ApJS..230...16K}. Again, we used VOSA to do a chi-square test to obtain the best-fit models and estimate spectral types for the objects. This library is created using spectra from the Sloan Digital Sky Survey's Baryon Oscillation Spectroscopic Survey (BOSS) and includes spectral templates covering spectral types from O5 through L3, they are binned by metallicity from $-$2.0 dex through $+$1.0 dex, and are separated into main-sequence dwarf stars and giants. Metallicity bins are extended down through the spectral sub-type M8, and the wavelength coverage is from 3650 to 10200 angstroms at a resolution of R~$\sim$~2000 \citep{2017ApJS..230...16K}.

\subsection{TSPEC/SOAR Spectra}

We obtained near-infrared (NIR) low-resolution spectra for two of the objects with the TripleSpec instrument mounted on the SOAR 4.1m telescope located at Cerro Pach\'on, in Chile. We observed in the usual ABBA pattern and obtained telluric standard A0V stars right after the observations of the source at a similar airmass. For the object \object{EDR3 5931456492784696064,} we obtained four 90 second exposures, and the telluric star was \object{HIP 81692};  for object \object{EDR3 6062219761378864128,} we obtained eight 120 second exposures, and the observed telluric was \object{HIP 64582}. 

For the data reduction for TripleSpec spectra, we used the IDL {\fontfamily{qcr}\selectfont spextools} package \citep{2004PASP..116..362C}. This includes wavelength calibration, flat fielding, sky subtraction, and extraction of the individual A and B positions. We then combined all the exposures for each object using the {\fontfamily{qcr}\selectfont xcombspec} task. We adopted a robust weighted mean to combine the spectra.
The final step was to correct the instrumental response function and telluric correction and merge the different orders. In order to do this, we used the {\fontfamily{qcr}\selectfont xtellcor} software \citep{2003PASP..115..389V} and {\fontfamily{qcr}\selectfont xmergeorders} task.

\begin{figure}[h]
\centering
%\hspace*{-0.3cm}
\includegraphics[width=\linewidth]{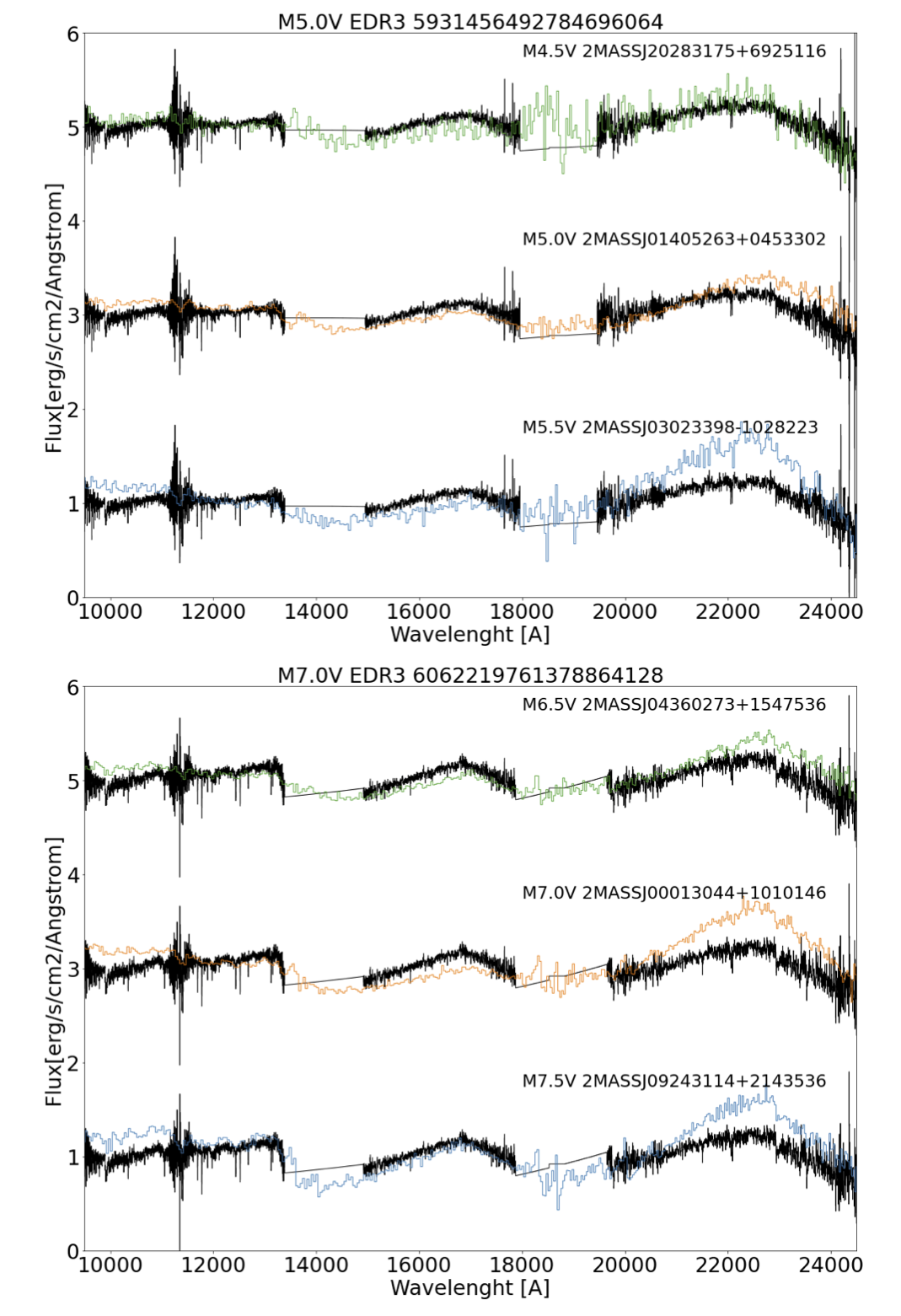}
\caption{Near-IR spectra for objects EDR3 5931456492784696064 and EDR3 6062219761378864128. Bad regions were removed with the {\fontfamily{qcr}\selectfont xcleanspex} IDL task. Each object is superimposed on different spectral templates from the SpeX Prism Library to compare the results obtained with VOSA. The identifier of each spectral template is indicated above each of them.}
\label{spectrum}
\end{figure}

In Figure \ref{spectrum}, we show the spectra compared with spectral templates of $\pm$ 1 sub-spectral type in 0.5 spectral type steps. These templates were collected from the public SpeX Prism Library\footnote{http://pono.ucsd.edu/~adam/browndwarfs/spexprism/} and come from \citet{2004AJ....127.2856B} and \citet{2010ApJS..190..100K}. The spectral type of the best-fit spectrum was assigned to our objects and then we derived the effective temperature from the stellar color and effective temperature sequence from \citet{2013ApJS..208....9P}.

\section{Results}
\label{results}

\begin{figure}[h]
\centering
%\hspace*{-1.0cm}
%\vspace*{2.0cm}
\includegraphics[width=\linewidth]{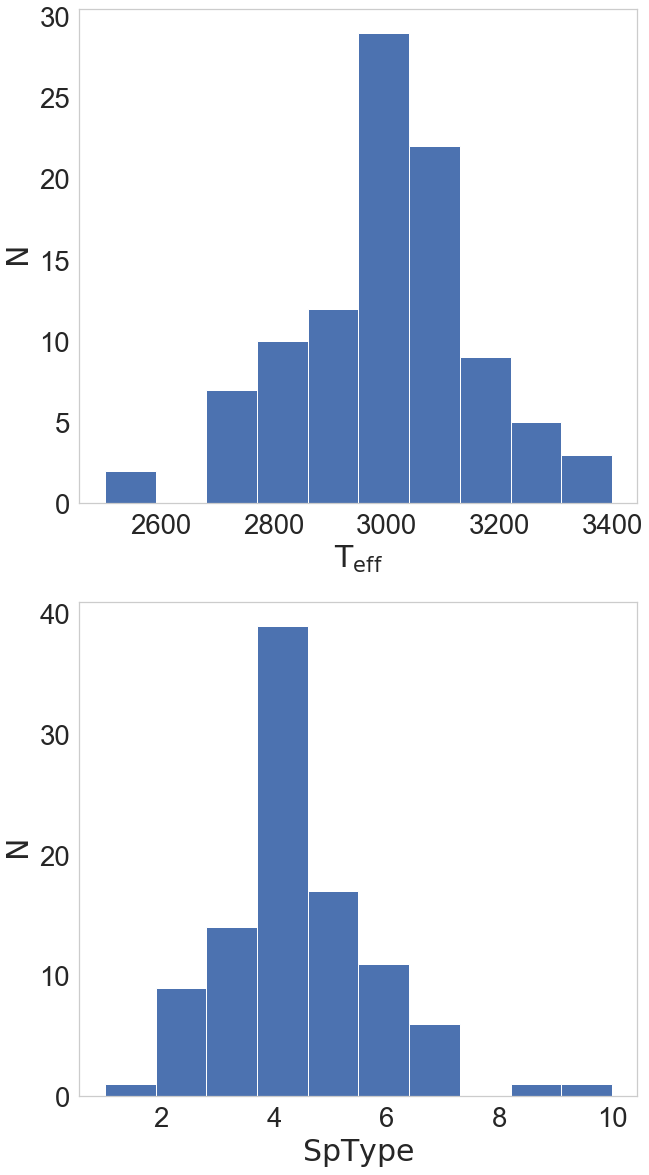}
\caption{Distributions of effective temperatures (top panel) and spectral types (lower panel) of the 99 characterized objects with VOSA. The mean values for effective temperature and spectral types are 3004 K and 4.3, respectively.}
\label{hists}
\end{figure}

The stellar parameters were computed using VOSA (see Section \ref{param}). The characterized sample is composed of a majority of M4 stars ($\bar{x} = 4.3$ and $\sigma = 1.5$), with a mean effective temperature of $\sim 3000$~K and a standard deviation of $\sim 178.61$~K. Since the BT-Settl theoretical model is fixed at solar metallicity, our sample is assumed to have $[Fe/H] = 0.0$ dex.  The distributions of effective temperatures and spectral types are shown in Fig. \ref{hists}.

We obtained a mean value for the mass of 0.11 $M_{\odot}$, with a minimum of 0.075 $M_{\odot}$ and a maximum of 0.27 $M_{\odot}$ for 61 objects of our sample. With respect to age, we report only one object (EDR3 5352517097624883584) with a measured age of 4.87$^{+6.67}$ Gyr, and three more objects with upper limits of 3.0, 1.82, and 2.84 Gyr.  

\begin{figure}[h!]
\centering
%\hspace*{-0.4cm}
\includegraphics[width=\linewidth]{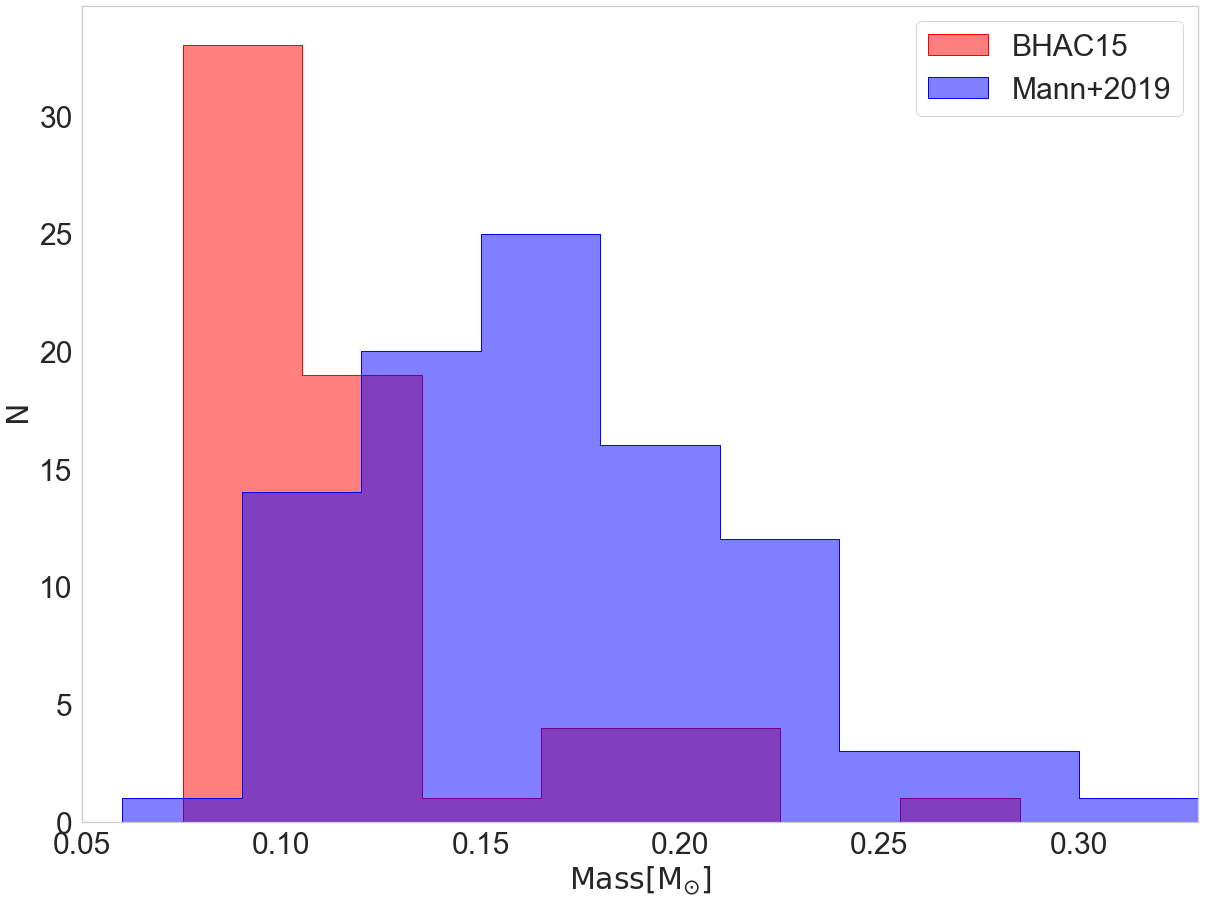}
\caption{Distribution of mass values obtained from evolutionary tracks of \citet{2015A&A...577A..42B} (in red) and from the $M_{Ks} - M_{*}$ relation of \citet{2019ApJ...871...63M} (in blue). VOSA results are observed to be more concentrated towards lower masses.}
\label{mass}
\end{figure}

We also used the $M_{Ks} - M_{*}$ relation from the work of \citet{2019ApJ...871...63M}, and obtained masses for 95 objects, since the remaining four did not lie between the limits of the relation. The results are similar to those computed with VOSA ($\bar{x_{Mass}} = 0.16 M_{\odot}$ ; $Mass_{min} = 0.085 M_{\odot}$ ; $Mass_{max} = 0.31 M_{\odot}$), but the distribution is different, given that VOSA results are much more concentrated towards lower masses (see Fig. \ref{mass}).  

\subsection{Comparison between stellar parameters obtained with VOSA and TSPEC spectra}

From Fig. \ref{spectrum}, we selected the best-fit spectral template for our objects and assigned spectral types to then estimate the effective temperatures with the stellar color and effective temperature sequence from \citet{2013ApJS..208....9P}. We determined the best fit for both objects by comparing the spectral template with the observed spectra by eye, focusing on the K$_{s}$ band.

For the first star plotted in Fig. \ref{spectrum}, VOSA assigned a spectral type of M5.0V, with an effective temperature of 3000 $\pm$ 50 K. From the spectral templates, we observed that the best fit is obtained with the M4.5V (shown in red). This indicates an effective temperature around 3060 K, which is in good agreement with the temperature estimated from VOSA. 

For our other object, we obtained a spectral type of M7.0V with VOSA, but we assigned a spectral type of M6.5V following the spectral template that best fits the spectrum (2MASS J04360273+1547536 in red). Again, from the \citet{2013ApJS..208....9P} sequence, we can estimate an effective temperature for this object of 2740 K, which is in good agreement with the temperature estimated with VOSA which is 2800 $\pm$ 50 K. 

\subsection{Comparison with an independent sample from literature}

To test our results, we sought a complete independent sample of stars with similar spectral types and similar photometry that also have computed stellar parameters. We found 11 stars to form the basis of a comparison, drawn from the works of \citet{2019AJ....158...56H} and \citet{2019ApJ...878..134K}, of which 6 have stellar parameters computed in both works.

\begin{figure}[h!]
\centering
%\hspace*{-0.5cm}
\includegraphics[width=\linewidth]{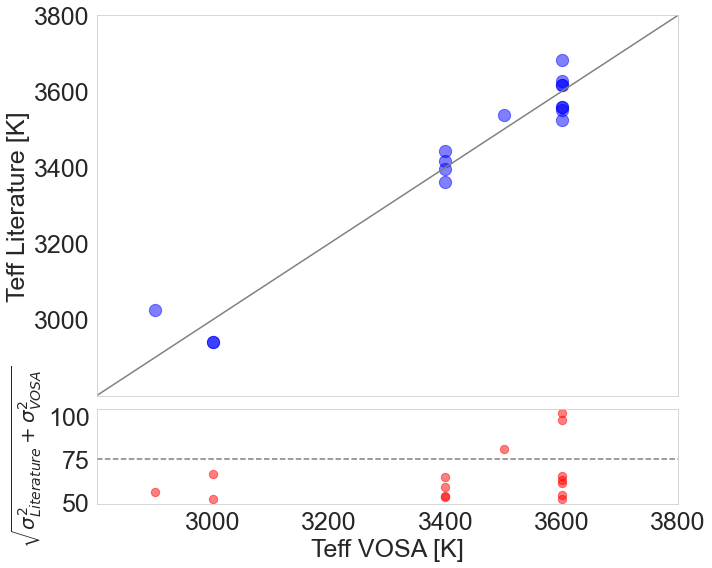}
\caption{Effective temperatures for stars in the works of \citet{2019AJ....158...56H} and \citet{2019ApJ...878..134K} compared to our values after computing them with our procedure (top panel). In gray, the line  represents a one-to-one relation. All the stars can be found near this one-to-one relation, indicating a good agreement between these values. In the bottom panel, combined errors are shown together with a dashed line that indicates the value for 3$\sigma$.}
\label{literature}
\end{figure}

We applied our procedure (described in Sects. \ref{seccion41} and \ref{seccion42}) to again compute the stellar parameters for these stars. Our estimated values and combined errors are in good agreement, with all the latter within 3-sigma except for three stars. Those have the highest values of error reported in the literature and, in particular, the star that is furthest from 3-sigma has stellar parameters calculated in both papers, when used as a comparison, only shows this high value of combined error when considering the error reported in \citet{2019ApJ...878..134K}, which is higher than the one in \citet{2019AJ....158...56H} (89 K and 38 K, respectively). Furthermore, we note that there is no clear trend that our method over- or underestimates the values for effective temperatures,  since objects can be found both above and below the one-to-one relation marked as a gray line in Fig. \ref{literature}.

\begin{figure}[h]
\centering
%\hspace*{-0.9cm}
\includegraphics[width=\linewidth]{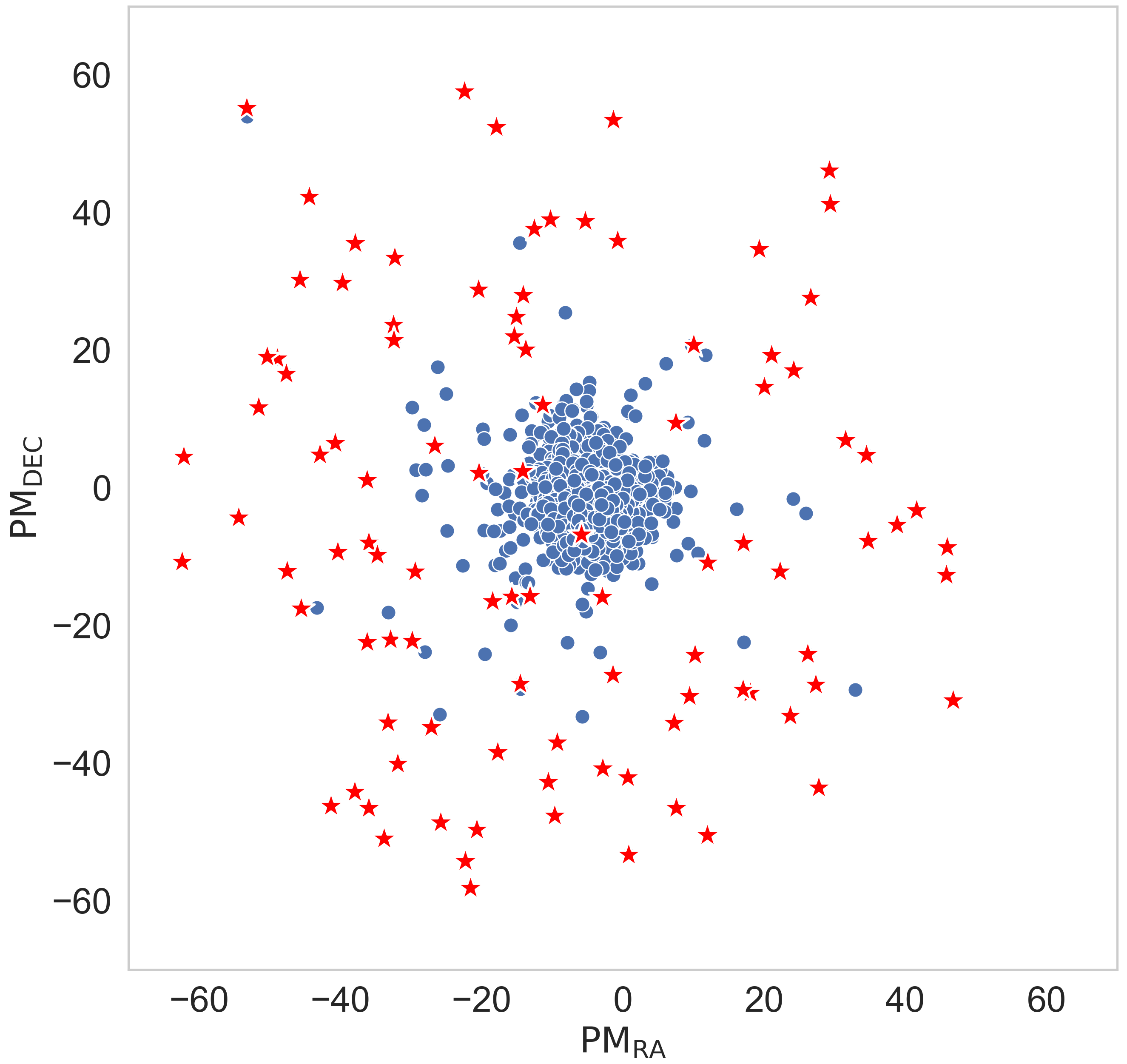}
\caption{Proper motion diagram for the 99 sources detected in the VVVX footprint (red stars) with respect to the proper motions in right ascension and declination coordinates of nearby objects within 30'' (blue dots).}
\label{pm_diag}
\end{figure}

\section{Discussion}
\label{discussion}

\begin{figure*}[t]
    \centering
    %\hspace*{-0.7cm}
    \includegraphics[width=\textwidth]{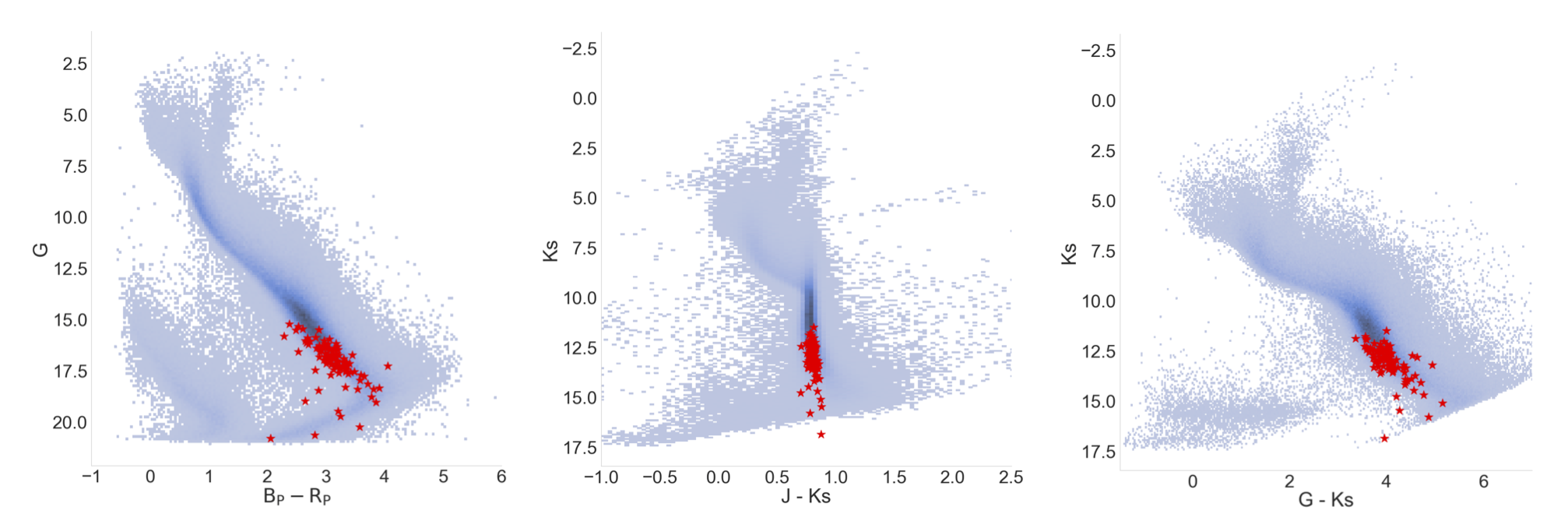}
    \caption{ Color-magnitude diagrams of our sample of nearby red-dwarf candidates (in red) compared with the entire \emph{Gaia} Catalog of Nearby Stars shown in blue. From left to right, $G$ vs ($B_{P} - R_{P}$), $K_{s}$ vs ($J - K_{s}$) and $K_{s}$ vs ($G - K_{s}$) diagrams are shown. In the three plots, our objects are located at the faintest magnitudes, and are mainly clustered in the main sequence.}
    \label{cmds_gcns}
\end{figure*}

Color-magnitude diagrams (CMDs) with different combinations of photometric bands for the objects in this sample and the entire GCNS are presented in Fig. \ref{cmds_gcns}. In the left panel, we present the CMD for the $G$, $B_{P}$ and $R_{P}$ bands of \emph{Gaia}. The white dwarf sequence is located at the bottom left part, at a bluer ($B_{P} - R_{P}$) color and fainter magnitudes than the rest of the GCNS sample (0.0 $\lesssim$ ($B_{P} - R_{P}$) $\lesssim$ 2.0 ; $G$ $>$ 12.5). The core H burning main sequence stars lie in the darkest region of the CMD, and our stars are concentrated in this region.

Some objects are shown at the faint end of the CMD with bluer ($B_{P} - R_{P}$) color (2.0 $\lesssim$ ($B_{P} - R_{P}$) $\lesssim$ 4.0 ; $G$ $>$ 19.0). It is important to note that, as mentioned in \citet{2019MNRAS.485.4423S}, the \emph{Gaia} $B_{P}$ band is intrinsically imprecise due to the very low flux of these objects at this wavelength range. We were not able to compute masses for these objects with the $M_{Ks} - M_{*}$ relation since it is only valid for red dwarfs between 4.0 < M$_{K_{s}}$ < 11.0. Only one has a mass estimated with VOSA which is 0.079$M_{\odot}$, with spectral type L0. On the other hand, at redder colors above the main sequence, there is the object \object{ERD3 6062219761378864128}, for which we did not identify any signs of binarity when probing its astrometric information and its RUWE. We observed this object with TSPEC in the near-infrared and obtained a spectral type of M7.0 with 2800 K. 

The central panel of Fig. \ref{cmds_gcns} shows the near-infrared CMD. The white dwarf sequence is not visible as these objects are too faint in these bands. The main sequence is shown around J - Ks $\sim$ 0.0 for Ks > 7.5 and our objects are located at the faintest end. Finally, in the right panel, we show G-Ks versus Ks CMD. Again, it is possible to distinguish the white dwarf sequence at K$_{s}$ > 12.5 and G - K$_{s}$ > 3.0, and the main sequence at K$_{s}$ > 7.5 and 1.0 < G - K$_{s}$ < 5.0 . Our stars are observed more clustered in the main sequence.

Wide separation binaries can be identified as pairs of stars with similar distances and matching proper motions. In Fig. \ref{pm_diag}, we present the proper-motion diagram for the sources included in this sample. We obtained proper motion values from the \emph{Gaia} EDR3 archive for all sources included within a circle of $30''$ around each one of our targets. We checked the proper motion vectors of our objects in the DECaPS DR1 color image composed with the \emph{g, r,} and \emph{z} bands, and compared these with the proper motions of the sources included in a circle of radius of $30''$. We classified seven objects of our sample as possible members of binary systems since their astrometric information (parallax and proper motion) was compatible with the ones of a nearby source. In particular, the object \object{EDR3 5693736443790365952} is found along with a brown dwarf candidate companion previously known from \citet{2018A&A...619L...8R}. We flagged all these objects as {\fontfamily{qcr}\selectfont BIN} in Table 4, where we present our complete results for the stellar parameters of the 99 objects of the sample. For five of our {\fontfamily{qcr}\selectfont BIN} candidates, the co-moving companion is not included in our sample of low-mass stars, but is present in the GCNS -- except one that is neither found in our sample nor in the GCNS. The estimated distance for this system is a little bit larger than 100 pc (103.6 pc), and since the objects included in the \emph{Gaia} Catalog of Nearby Stars were only required to have a non-zero probability to be within 100 pc, it is possible that only one of the components of the system was included in the catalog. We present their positions and \emph{Gaia} photometry in Table 5.

\begin{figure*}[t]
    \centering
    %\hspace*{0.0cm}
    \includegraphics[width=\textwidth]{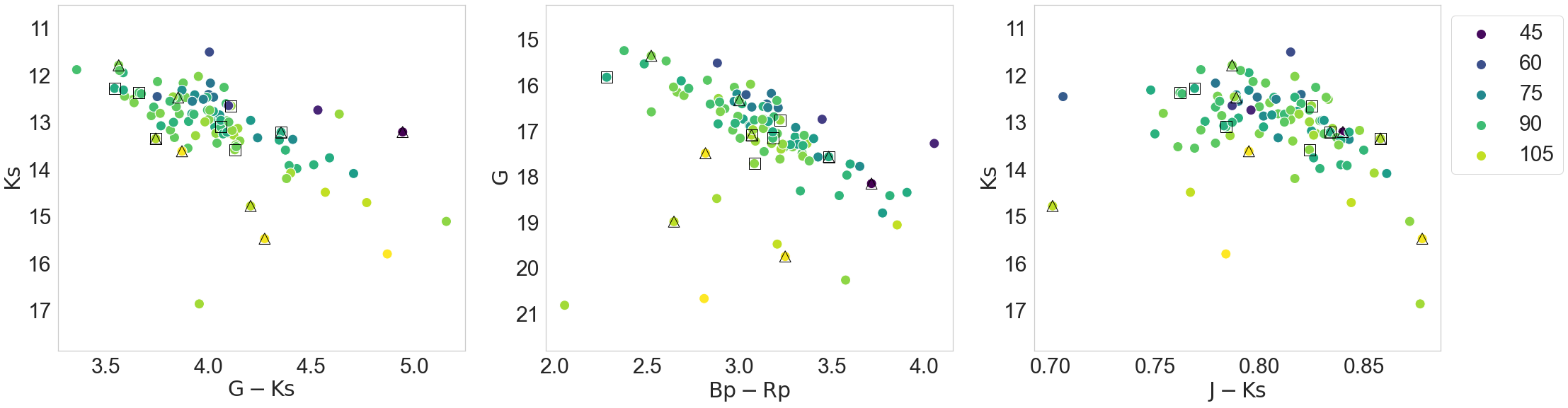}
    \caption{Zoom-in CMDs of our sample of nearby red-dwarf candidates from Fig. \ref{cmds_gcns}. For the three CMDs, unfilled black squares represent the objects flagged as {\fontfamily{qcr}\selectfont BIN} because of the similarities between their astrometric information and any other near object within 30''. Black unfilled triangles indicate objects flagged as {\fontfamily{qcr}\selectfont HR}, possibly containing unresolved companions. These CMDs are color-coded by the distance from the Sun, which is obtained from the Bayesian analysis of Gaia EDR3 parallaxes of \citet{2021AJ....161..147B}, with darker circles being closer.}
    \label{cmds_zoom}
\end{figure*}

For stars with unresolved companions, a single-source astrometric model performs poorly. Such stars can be easily detected with the reduced $\chi^2$ statistics, or the renormalized unit weight error \citep[RUWE;][]{2020MNRAS.496.1922B}. We defined that objects that are further from a single-source fitting from the astrometric solution are those with a RUWE $\geq$ 1.4. We marked them in Fig. \ref{cmds_zoom} as unfilled black triangles, and used the flag {\fontfamily{qcr}\selectfont HR} in Table 4. In our sample, six sources have a high RUWE value, implying the possibility of unresolved companions. In addition, we identified two objects that were previously classified as binary systems and present at the same time, a high RUWE value, pointing again to unresolved companions. We propose that these systems could be triple systems, with one component that is resolved and two more that could not be identified as individual objects. We marked them as ({\fontfamily{qcr}\selectfont Multiple system}) in Table 4.

We searched for our objects in the SIMBAD Astronomical Database \citep{2000A&AS..143....9W} and found 3 objects. The first one is \object{EDR3 5254655030851879040}, which is included in the work of \citet{2018ApJ...865..136G} as an M5 that forms part of the Volans-Carina stellar association, which is located at a distance of $\sim$ 75-100 pc with an estimated age of $\sim$ 90 Myr. The other two objects are \object{EDR3 5257810972787599360} and \object{EDR3 5409811755160786048}, and both are classified as brown dwarf candidates in the ultra-cool dwarf catalog from \emph{Gaia} DR2 data from \citet{2018A&A...619L...8R}.

In addition, we performed a search around 1'' of the coordinates of the objects in the Vizier database \citep{2000A&AS..143...23O}. We highlight 14 of our objects that were present in catalogs of wide binaries \citep{2020ApJS..246....4T, 2020AN....341..996M, 2020ApJS..247...66H}, ultra-cool dwarfs candidates in \emph{Gaia} \citep{2018A&A...619L...8R}, and large-amplitude variables in \emph{Gaia} \citep{2021A&A...648A..44M}, together with catalogs with ultraviolet photometry \citep{2014Ap&SS.354...97Y} and lensed \emph{Gaia} systems \citep{2019A&A...622A.165D}. This references are indicated in Table 4.

\section{Conclusions}
\label{conclusions}

We identified a sample of 99 nearby red dwarf stars within 100 pc from the Sun from the \emph{Gaia} EDR3 that are in the VVVX footprint along with the external VVVX fields of the Galactic disk. To properly characterize their physical properties and spectral types, we first collected (in addition to their near-infrared \emph{JHK$_s$} magnitudes) the optical photometry in the \emph{grizY} bands from the DECaPS and Pan-STARRS surveys and mid-IR data from the WISE Survey when available (see Table 1 for astrometric information and Tables 2 and 3 for photometry). 

The spectral energy distribution of these objects was constructed from photometry by the Virtual Observatory SED Analyzer (VOSA) and then compared with the BT-Settl CIFIST \citep{2011ASPC..448...91A} theoretical models to estimate their stellar parameters and with the empirical spectral library from \citet{2017ApJS..230...16K} to obtain their spectral types. We obtained low-resolution spectra with the TSPEC instrument mounted on the SOAR telescope for two objects of the sample and we estimated stellar parameters by comparing these spectra with other spectral templates within $\pm$ 1 sub-spectral type. Spectral types classified using spectral templates are, for both cases, 0.5 sub-spectral types earlier. Temperatures from the \citet{2013ApJS..208....9P} sequence are in agreement with the errors. 

We additionally tested our procedure with a complete independent sample obtained from the works of \citet{2019AJ....158...56H} and \citet{2019ApJ...878..134K} and compared the stellar parameters reported by them with ours for 11 objects with similar spectral types and photometry. Values for effective temperatures are in agreement, with a nearly one-to-one relation. Combined errors are also calculated and all are within 3-sigma, except for three stars that present the highest errors in the literature.

We also used the \citet{2015A&A...577A..42B} isochrones and evolutionary models to determine masses and ages. We only report the age of one object to be 4.87$^{+6.67}$ Gyr, along with three additional upper limits for the age of 3.0, 1.82, and 2.84. We also computed masses with the $M_{Ks} - M_{*}$ relation from \citet{2019ApJ...871...63M} and found that the distribution of masses with VOSA are more concentrated to the very low-mass regime, while results from the $M_{Ks} - M_{*}$ relation are more spread out in terms of mass. 

We revised the 99 sources in the images of DECaPS in Aladin, together with their astrometric information from \emph{Gaia} EDR3 with the aim to identify possible stellar or sub-stellar companions. Seven objects were classified as binaries based on their consistency with regard to the parallaxes and proper motions with nearby objects. Two of these binary systems also have RUWE values that are far from unity (RUWE $>$ 1.4), indicating an additional component that could be unresolved. The final stellar parameters are presented in Table 4.

\begin{acknowledgements}
We are very grateful to the anonymous referee for the useful comments and suggestions, which helped improve this paper. This work is based on data products from VVVX Survey observations made with the VISTA telescope at the ESO Paranal Observatory under program ID 198.B-2004. AM is funded by the National Agency for Research and Development (ANID) /  Scholarship Program / DOCTORADO BECAS CHILE / 2021 - 21210535. DM is supported by the BASAL Center for Astrophysics and Associated Technologies (CATA) through gran AFB-170002, and by FONDECYT Regular grant No. 1170121. J.A.-G. acknowledges support from FONDECYT Regular 1201490 and ANID - Millenium Science Iniciative Program - ICN12\_009 awarded to the Millenium Institue of Astrophysics MAS. JCB acknowledges support from FONDECYT postdoctorado 3180716. R.K.S. acknowledges support from CNPq/Brazil through project 305902/2019-9. This research has made use of the TOPCAT\footnote{http://www.starlink.ac.uk/topcat/} tool \citep{2005ASPC..347...29T}, VOSA\footnote{http://svo2.cab.inta-csic.es/theory/vosa/}, developed under the Spanish Virtual Observatory project supported by the Spanish MINECO through grant AyA2017-84089. This research has made use of "Aladin Sky Atlas" developed at CDS, Strasbourg Observatory, France. This research has also made use of the SIMBAD database, operated at CDS, Strasbourg, France. This work used data from the SpeX Prism Library service developed by the Spanish Virtual Observatory in the framework of the IAU Commission G5 Working Group: Spectral Stellar Libraries. This research has made use of the VizieR catalog access tool, CDS, Strasbourg, France (DOI:10.26093/cds/vizier). The original description 
 of the VizieR service was published in 2000, A\&AS 143, 23. The authors also acknowledge J.B. Pullen. 
\end{acknowledgements}

\bibliographystyle{aa} % style aa.bst
\bibliography{41759corr.bib} % your references Yourfile.bib

\begin{thebibliography}{62}
\expandafter\ifx\csname natexlab\endcsname\relax\def\natexlab#1{#1}\fi

\bibitem[{{Allard} {et~al.}(2011){Allard}, {Homeier}, \&
  {Freytag}}]{2011ASPC..448...91A}
{Allard}, F., {Homeier}, D., \& {Freytag}, B. 2011, in Astronomical Society of
  the Pacific Conference Series, Vol. 448, 16th Cambridge Workshop on Cool
  Stars, Stellar Systems, and the Sun, ed. C.~{Johns-Krull}, M.~K. {Browning},
  \& A.~A. {West}, 91

\bibitem[{{Alonso-Garc{\'\i}a} {et~al.}(2018){Alonso-Garc{\'\i}a}, {Saito},
  {Hempel}, {Minniti}, {Pullen}, {Catelan}, {Ramos}, {Cross}, {Gonzalez},
  {Lucas}, {Palma}, {Valenti}, \& {Zoccali}}]{2018A&A...619A...4A}
{Alonso-Garc{\'\i}a}, J., {Saito}, R.~K., {Hempel}, M., {et~al.} 2018, \aap,
  619, A4

\bibitem[{{Bailer-Jones} {et~al.}(2021){Bailer-Jones}, {Rybizki}, {Fouesneau},
  {Demleitner}, \& {Andrae}}]{2021AJ....161..147B}
{Bailer-Jones}, C.~A.~L., {Rybizki}, J., {Fouesneau}, M., {Demleitner}, M., \&
  {Andrae}, R. 2021, \aj, 161, 147

\bibitem[{{Baraffe} {et~al.}(2015){Baraffe}, {Homeier}, {Allard}, \&
  {Chabrier}}]{2015A&A...577A..42B}
{Baraffe}, I., {Homeier}, D., {Allard}, F., \& {Chabrier}, G. 2015, \aap, 577,
  A42

\bibitem[{{Bayo} {et~al.}(2008){Bayo}, {Rodrigo}, {Barrado Y Navascu{\'e}s},
  {Solano}, {Guti{\'e}rrez}, {Morales-Calder{\'o}n}, \&
  {Allard}}]{2008A&A...492..277B}
{Bayo}, A., {Rodrigo}, C., {Barrado Y Navascu{\'e}s}, D., {et~al.} 2008, \aap,
  492, 277

\bibitem[{{Beam{\'\i}n} {et~al.}(2018){Beam{\'\i}n}, {Mendez}, {Smart}, {Jara},
  {Kurtev}, {Gromadzki}, {Villanueva}, {Minniti}, {Smith}, \&
  {Lucas}}]{2018RMxAC..50...55B}
{Beam{\'\i}n}, J.~C., {Mendez}, R.~A., {Smart}, R.~L., {et~al.} 2018, in
  Revista Mexicana de Astronomia y Astrofisica Conference Series, Vol.~50,
  Revista Mexicana de Astronomia y Astrofisica Conference Series, 55--55

\bibitem[{{Beam{\'\i}n} {et~al.}(2013){Beam{\'\i}n}, {Minniti}, {Gromadzki},
  {Kurtev}, {Ivanov}, {Beletsky}, {Lucas}, {Saito}, \&
  {Borissova}}]{2013A&A...557L...8B}
{Beam{\'\i}n}, J.~C., {Minniti}, D., {Gromadzki}, M., {et~al.} 2013, \aap, 557,
  L8

\bibitem[{{Belokurov} {et~al.}(2020){Belokurov}, {Penoyre}, {Oh}, {Iorio},
  {Hodgkin}, {Evans}, {Everall}, {Koposov}, {Tout}, {Izzard}, {Clarke}, \&
  {Brown}}]{2020MNRAS.496.1922B}
{Belokurov}, V., {Penoyre}, Z., {Oh}, S., {et~al.} 2020, \mnras, 496, 1922

\bibitem[{{Burgasser} {et~al.}(2004){Burgasser}, {McElwain}, {Kirkpatrick},
  {Cruz}, {Tinney}, \& {Reid}}]{2004AJ....127.2856B}
{Burgasser}, A.~J., {McElwain}, M.~W., {Kirkpatrick}, J.~D., {et~al.} 2004,
  \aj, 127, 2856

\bibitem[{{Casagrande} {et~al.}(2011){Casagrande}, {Sch{\"o}nrich}, {Asplund},
  {Cassisi}, {Ram{\'\i}rez}, {Mel{\'e}ndez}, {Bensby}, \&
  {Feltzing}}]{2011A&A...530A.138C}
{Casagrande}, L., {Sch{\"o}nrich}, R., {Asplund}, M., {et~al.} 2011, \aap, 530,
  A138

\bibitem[{{Chambers} {et~al.}(2016){Chambers}, {Magnier}, {Metcalfe},
  {Flewelling}, {Huber}, {Waters}, {Denneau}, {Draper}, {Farrow}, {Finkbeiner},
  {Holmberg}, {Koppenhoefer}, {Price}, {Rest}, {Saglia}, {Schlafly}, {Smartt},
  {Sweeney}, {Wainscoat}, {Burgett}, {Chastel}, {Grav}, {Heasley}, {Hodapp},
  {Jedicke}, {Kaiser}, {Kudritzki}, {Luppino}, {Lupton}, {Monet}, {Morgan},
  {Onaka}, {Shiao}, {Stubbs}, {Tonry}, {White}, {Ba{\~n}ados}, {Bell},
  {Bender}, {Bernard}, {Boegner}, {Boffi}, {Botticella}, {Calamida},
  {Casertano}, {Chen}, {Chen}, {Cole}, {Deacon}, {Frenk}, {Fitzsimmons},
  {Gezari}, {Gibbs}, {Goessl}, {Goggia}, {Gourgue}, {Goldman}, {Grant},
  {Grebel}, {Hambly}, {Hasinger}, {Heavens}, {Heckman}, {Henderson}, {Henning},
  {Holman}, {Hopp}, {Ip}, {Isani}, {Jackson}, {Keyes}, {Koekemoer}, {Kotak},
  {Le}, {Liska}, {Long}, {Lucey}, {Liu}, {Martin}, {Masci}, {McLean}, {Mindel},
  {Misra}, {Morganson}, {Murphy}, {Obaika}, {Narayan}, {Nieto-Santisteban},
  {Norberg}, {Peacock}, {Pier}, {Postman}, {Primak}, {Rae}, {Rai}, {Riess},
  {Riffeser}, {Rix}, {R{\"o}ser}, {Russel}, {Rutz}, {Schilbach}, {Schultz},
  {Scolnic}, {Strolger}, {Szalay}, {Seitz}, {Small}, {Smith}, {Soderblom},
  {Taylor}, {Thomson}, {Taylor}, {Thakar}, {Thiel}, {Thilker}, {Unger},
  {Urata}, {Valenti}, {Wagner}, {Walder}, {Walter}, {Watters}, {Werner},
  {Wood-Vasey}, \& {Wyse}}]{2016arXiv161205560C}
{Chambers}, K.~C., {Magnier}, E.~A., {Metcalfe}, N., {et~al.} 2016, arXiv
  e-prints, arXiv:1612.05560

\bibitem[{{Cushing} {et~al.}(2004){Cushing}, {Vacca}, \&
  {Rayner}}]{2004PASP..116..362C}
{Cushing}, M.~C., {Vacca}, W.~D., \& {Rayner}, J.~T. 2004, \pasp, 116, 362

\bibitem[{{Delchambre} {et~al.}(2019){Delchambre}, {Krone-Martins}, {Wertz},
  {Ducourant}, {Galluccio}, {Kl{\"u}ter}, {Mignard}, {Teixeira}, {Djorgovski},
  {Stern}, {Graham}, {Surdej}, {Bastian}, {Wambsganss}, {Le Campion}, \&
  {Slezak}}]{2019A&A...622A.165D}
{Delchambre}, L., {Krone-Martins}, A., {Wertz}, O., {et~al.} 2019, \aap, 622,
  A165

\bibitem[{{Dole}(1964)}]{1964hpfm.book.....D}
{Dole}, S.~H. 1964, {Habitable planets for man}

\bibitem[{{Dressing} \& {Charbonneau}(2015)}]{2015ApJ...807...45D}
{Dressing}, C.~D. \& {Charbonneau}, D. 2015, \apj, 807, 45

\bibitem[{{Feliz} {et~al.}(2021){Feliz}, {Plavchan}, {Bianco}, {Jimenez},
  {Collins}, {Villarreal Alvarado}, \& {Stassun}}]{2021AJ....161..247F}
{Feliz}, D.~L., {Plavchan}, P., {Bianco}, S.~N., {et~al.} 2021, \aj, 161, 247

\bibitem[{{Gagn{\'e}} {et~al.}(2018){Gagn{\'e}}, {Faherty}, \&
  {Mamajek}}]{2018ApJ...865..136G}
{Gagn{\'e}}, J., {Faherty}, J.~K., \& {Mamajek}, E.~E. 2018, \apj, 865, 136

\bibitem[{{Gaia Collaboration} {et~al.}(2018){Gaia Collaboration}, {Brown},
  {Vallenari}, {Prusti}, {de Bruijne}, {Babusiaux}, {Bailer-Jones}, {Biermann},
  {Evans}, {Eyer}, {Jansen}, {Jordi}, {Klioner}, {Lammers}, {Lindegren},
  {Luri}, {Mignard}, {Panem}, {Pourbaix}, {Randich}, {Sartoretti}, {Siddiqui},
  {Soubiran}, {van Leeuwen}, {Walton}, {Arenou}, {Bastian}, {Cropper},
  {Drimmel}, {Katz}, {Lattanzi}, {Bakker}, {Cacciari}, {Casta{\~n}eda},
  {Chaoul}, {Cheek}, {De Angeli}, {Fabricius}, {Guerra}, {Holl}, {Masana},
  {Messineo}, {Mowlavi}, {Nienartowicz}, {Panuzzo}, {Portell}, {Riello},
  {Seabroke}, {Tanga}, {Th{\'e}venin}, {Gracia-Abril}, {Comoretto},
  {Garcia-Reinaldos}, {Teyssier}, {Altmann}, {Andrae}, {Audard},
  {Bellas-Velidis}, {Benson}, {Berthier}, {Blomme}, {Burgess}, {Busso},
  {Carry}, {Cellino}, {Clementini}, {Clotet}, {Creevey}, {Davidson}, {De
  Ridder}, {Delchambre}, {Dell'Oro}, {Ducourant},
  {Fern{\'a}ndez-Hern{\'a}ndez}, {Fouesneau}, {Fr{\'e}mat}, {Galluccio},
  {Garc{\'\i}a-Torres}, {Gonz{\'a}lez-N{\'u}{\~n}ez}, {Gonz{\'a}lez-Vidal},
  {Gosset}, {Guy}, {Halbwachs}, {Hambly}, {Harrison}, {Hern{\'a}ndez},
  {Hestroffer}, {Hodgkin}, {Hutton}, {Jasniewicz}, {Jean-Antoine-Piccolo},
  {Jordan}, {Korn}, {Krone-Martins}, {Lanzafame}, {Lebzelter}, {L{\"o}ffler},
  {Manteiga}, {Marrese}, {Mart{\'\i}n-Fleitas}, {Moitinho}, {Mora}, {Muinonen},
  {Osinde}, {Pancino}, {Pauwels}, {Petit}, {Recio-Blanco}, {Richards},
  {Rimoldini}, {Robin}, {Sarro}, {Siopis}, {Smith}, {Sozzetti}, {S{\"u}veges},
  {Torra}, {van Reeven}, {Abbas}, {Abreu Aramburu}, {Accart}, {Aerts},
  {Altavilla}, {{\'A}lvarez}, {Alvarez}, {Alves}, {Anderson}, {Andrei},
  {Anglada Varela}, {Antiche}, {Antoja}, {Arcay}, {Astraatmadja}, {Bach},
  {Baker}, {Balaguer-N{\'u}{\~n}ez}, {Balm}, {Barache}, {Barata}, {Barbato},
  {Barblan}, {Barklem}, {Barrado}, {Barros}, {Barstow}, {Bartholom{\'e}
  Mu{\~n}oz}, {Bassilana}, {Becciani}, {Bellazzini}, {Berihuete}, {Bertone},
  {Bianchi}, {Bienaym{\'e}}, {Blanco-Cuaresma}, {Boch}, {Boeche}, {Bombrun},
  {Borrachero}, {Bossini}, {Bouquillon}, {Bourda}, {Bragaglia}, {Bramante},
  {Breddels}, {Bressan}, {Brouillet}, {Br{\"u}semeister}, {Brugaletta},
  {Bucciarelli}, {Burlacu}, {Busonero}, {Butkevich}, {Buzzi}, {Caffau},
  {Cancelliere}, {Cannizzaro}, {Cantat-Gaudin}, {Carballo}, {Carlucci},
  {Carrasco}, {Casamiquela}, {Castellani}, {Castro-Ginard}, {Charlot},
  {Chemin}, {Chiavassa}, {Cocozza}, {Costigan}, {Cowell}, {Crifo}, {Crosta},
  {Crowley}, {Cuypers}, {Dafonte}, {Damerdji}, {Dapergolas}, {David}, {David},
  {de Laverny}, {De Luise}, {De March}, {de Martino}, {de Souza}, {de Torres},
  {Debosscher}, {del Pozo}, {Delbo}, {Delgado}, {Delgado}, {Di Matteo},
  {Diakite}, {Diener}, {Distefano}, {Dolding}, {Drazinos}, {Dur{\'a}n},
  {Edvardsson}, {Enke}, {Eriksson}, {Esquej}, {Eynard Bontemps}, {Fabre},
  {Fabrizio}, {Faigler}, {Falc{\~a}o}, {Farr{\`a}s Casas}, {Federici},
  {Fedorets}, {Fernique}, {Figueras}, {Filippi}, {Findeisen}, {Fonti},
  {Fraile}, {Fraser}, {Fr{\'e}zouls}, {Gai}, {Galleti}, {Garabato},
  {Garc{\'\i}a-Sedano}, {Garofalo}, {Garralda}, {Gavel}, {Gavras}, {Gerssen},
  {Geyer}, {Giacobbe}, {Gilmore}, {Girona}, {Giuffrida}, {Glass}, {Gomes},
  {Granvik}, {Gueguen}, {Guerrier}, {Guiraud}, {Guti{\'e}rrez-S{\'a}nchez},
  {Haigron}, {Hatzidimitriou}, {Hauser}, {Haywood}, {Heiter}, {Helmi}, {Heu},
  {Hilger}, {Hobbs}, {Hofmann}, {Holland}, {Huckle}, {Hypki}, {Icardi},
  {Jan{\ss}en}, {Jevardat de Fombelle}, {Jonker}, {Juh{\'a}sz}, {Julbe},
  {Karampelas}, {Kewley}, {Klar}, {Kochoska}, {Kohley}, {Kolenberg},
  {Kontizas}, {Kontizas}, {Koposov}, {Kordopatis}, {Kostrzewa-Rutkowska},
  {Koubsky}, {Lambert}, {Lanza}, {Lasne}, {Lavigne}, {Le Fustec}, {Le
  Poncin-Lafitte}, {Lebreton}, {Leccia}, {Leclerc}, {Lecoeur-Taibi},
  {Lenhardt}, {Leroux}, {Liao}, {Licata}, {Lindstr{\o}m}, {Lister}, {Livanou},
  {Lobel}, {L{\'o}pez}, {Managau}, {Mann}, {Mantelet}, {Marchal}, {Marchant},
  {Marconi}, {Marinoni}, {Marschalk{\'o}}, {Marshall}, {Martino}, {Marton},
  {Mary}, {Massari}, {Matijevi{\v{c}}}, {Mazeh}, {McMillan}, {Messina},
  {Michalik}, {Millar}, {Molina}, {Molinaro}, {Moln{\'a}r}, {Montegriffo},
  {Mor}, {Morbidelli}, {Morel}, {Morris}, {Mulone}, {Muraveva}, {Musella},
  {Nelemans}, {Nicastro}, {Noval}, {O'Mullane}, {Ord{\'e}novic},
  {Ord{\'o}{\~n}ez-Blanco}, {Osborne}, {Pagani}, {Pagano}, {Pailler},
  {Palacin}, {Palaversa}, {Panahi}, {Pawlak}, {Piersimoni}, {Pineau}, {Plachy},
  {Plum}, {Poggio}, {Poujoulet}, {Pr{\v{s}}a}, {Pulone}, {Racero}, {Ragaini},
  {Rambaux}, {Ramos-Lerate}, {Regibo}, {Reyl{\'e}}, {Riclet}, {Ripepi}, {Riva},
  {Rivard}, {Rixon}, {Roegiers}, {Roelens}, {Romero-G{\'o}mez}, {Rowell},
  {Royer}, {Ruiz-Dern}, {Sadowski}, {Sagrist{\`a} Sell{\'e}s}, {Sahlmann},
  {Salgado}, {Salguero}, {Sanna}, {Santana-Ros}, {Sarasso}, {Savietto},
  {Schultheis}, {Sciacca}, {Segol}, {Segovia}, {S{\'e}gransan}, {Shih},
  {Siltala}, {Silva}, {Smart}, {Smith}, {Solano}, {Solitro}, {Sordo}, {Soria
  Nieto}, {Souchay}, {Spagna}, {Spoto}, {Stampa}, {Steele},
  {Steidelm{\"u}ller}, {Stephenson}, {Stoev}, {Suess}, {Surdej}, {Szabados},
  {Szegedi-Elek}, {Tapiador}, {Taris}, {Tauran}, {Taylor}, {Teixeira},
  {Terrett}, {Teyssandier}, {Thuillot}, {Titarenko}, {Torra Clotet}, {Turon},
  {Ulla}, {Utrilla}, {Uzzi}, {Vaillant}, {Valentini}, {Valette}, {van Elteren},
  {Van Hemelryck}, {van Leeuwen}, {Vaschetto}, {Vecchiato}, {Veljanoski},
  {Viala}, {Vicente}, {Vogt}, {von Essen}, {Voss}, {Votruba}, {Voutsinas},
  {Walmsley}, {Weiler}, {Wertz}, {Wevers}, {Wyrzykowski}, {Yoldas},
  {{\v{Z}}erjal}, {Ziaeepour}, {Zorec}, {Zschocke}, {Zucker}, {Zurbach}, \&
  {Zwitter}}]{2018A&A...616A...1G}
{Gaia Collaboration}, {Brown}, A.~G.~A., {Vallenari}, A., {et~al.} 2018, \aap,
  616, A1

\bibitem[{{Gaia Collaboration} {et~al.}(2021{\natexlab{a}}){Gaia
  Collaboration}, {Brown}, {Vallenari}, {Prusti}, {de Bruijne}, {Babusiaux},
  {Biermann}, {Creevey}, {Evans}, {Eyer}, {Hutton}, {Jansen}, {Jordi},
  {Klioner}, {Lammers}, {Lindegren}, {Luri}, {Mignard}, {Panem}, {Pourbaix},
  {Randich}, {Sartoretti}, {Soubiran}, {Walton}, {Arenou}, {Bailer-Jones},
  {Bastian}, {Cropper}, {Drimmel}, {Katz}, {Lattanzi}, {van Leeuwen}, {Bakker},
  {Cacciari}, {Casta{\~n}eda}, {De Angeli}, {Ducourant}, {Fabricius},
  {Fouesneau}, {Fr{\'e}mat}, {Guerra}, {Guerrier}, {Guiraud}, {Jean-Antoine
  Piccolo}, {Masana}, {Messineo}, {Mowlavi}, {Nicolas}, {Nienartowicz},
  {Pailler}, {Panuzzo}, {Riclet}, {Roux}, {Seabroke}, {Sordo}, {Tanga},
  {Th{\'e}venin}, {Gracia-Abril}, {Portell}, {Teyssier}, {Altmann}, {Andrae},
  {Bellas-Velidis}, {Benson}, {Berthier}, {Blomme}, {Brugaletta}, {Burgess},
  {Busso}, {Carry}, {Cellino}, {Cheek}, {Clementini}, {Damerdji}, {Davidson},
  {Delchambre}, {Dell'Oro}, {Fern{\'a}ndez-Hern{\'a}ndez}, {Galluccio},
  {Garc{\'\i}a-Lario}, {Garcia-Reinaldos}, {Gonz{\'a}lez-N{\'u}{\~n}ez},
  {Gosset}, {Haigron}, {Halbwachs}, {Hambly}, {Harrison}, {Hatzidimitriou},
  {Heiter}, {Hern{\'a}ndez}, {Hestroffer}, {Hodgkin}, {Holl}, {Jan{\ss}en},
  {Jevardat de Fombelle}, {Jordan}, {Krone-Martins}, {Lanzafame},
  {L{\"o}ffler}, {Lorca}, {Manteiga}, {Marchal}, {Marrese}, {Moitinho}, {Mora},
  {Muinonen}, {Osborne}, {Pancino}, {Pauwels}, {Petit}, {Recio-Blanco},
  {Richards}, {Riello}, {Rimoldini}, {Robin}, {Roegiers}, {Rybizki}, {Sarro},
  {Siopis}, {Smith}, {Sozzetti}, {Ulla}, {Utrilla}, {van Leeuwen}, {van
  Reeven}, {Abbas}, {Abreu Aramburu}, {Accart}, {Aerts}, {Aguado}, {Ajaj},
  {Altavilla}, {{\'A}lvarez}, {{\'A}lvarez Cid-Fuentes}, {Alves}, {Anderson},
  {Anglada Varela}, {Antoja}, {Audard}, {Baines}, {Baker},
  {Balaguer-N{\'u}{\~n}ez}, {Balbinot}, {Balog}, {Barache}, {Barbato},
  {Barros}, {Barstow}, {Bartolom{\'e}}, {Bassilana}, {Bauchet},
  {Baudesson-Stella}, {Becciani}, {Bellazzini}, {Bernet}, {Bertone}, {Bianchi},
  {Blanco-Cuaresma}, {Boch}, {Bombrun}, {Bossini}, {Bouquillon}, {Bragaglia},
  {Bramante}, {Breedt}, {Bressan}, {Brouillet}, {Bucciarelli}, {Burlacu},
  {Busonero}, {Butkevich}, {Buzzi}, {Caffau}, {Cancelliere}, {C{\'a}novas},
  {Cantat-Gaudin}, {Carballo}, {Carlucci}, {Carnerero}, {Carrasco},
  {Casamiquela}, {Castellani}, {Castro-Ginard}, {Castro Sampol}, {Chaoul},
  {Charlot}, {Chemin}, {Chiavassa}, {Cioni}, {Comoretto}, {Cooper}, {Cornez},
  {Cowell}, {Crifo}, {Crosta}, {Crowley}, {Dafonte}, {Dapergolas}, {David},
  {David}, {de Laverny}, {De Luise}, {De March}, {De Ridder}, {de Souza}, {de
  Teodoro}, {de Torres}, {del Peloso}, {del Pozo}, {Delbo}, {Delgado},
  {Delgado}, {Delisle}, {Di Matteo}, {Diakite}, {Diener}, {Distefano},
  {Dolding}, {Eappachen}, {Edvardsson}, {Enke}, {Esquej}, {Fabre}, {Fabrizio},
  {Faigler}, {Fedorets}, {Fernique}, {Fienga}, {Figueras}, {Fouron},
  {Fragkoudi}, {Fraile}, {Franke}, {Gai}, {Garabato}, {Garcia-Gutierrez},
  {Garc{\'\i}a-Torres}, {Garofalo}, {Gavras}, {Gerlach}, {Geyer}, {Giacobbe},
  {Gilmore}, {Girona}, {Giuffrida}, {Gomel}, {Gomez}, {Gonzalez-Santamaria},
  {Gonz{\'a}lez-Vidal}, {Granvik}, {Guti{\'e}rrez-S{\'a}nchez}, {Guy},
  {Hauser}, {Haywood}, {Helmi}, {Hidalgo}, {Hilger}, {H{\l}adczuk}, {Hobbs},
  {Holland}, {Huckle}, {Jasniewicz}, {Jonker}, {Juaristi Campillo}, {Julbe},
  {Karbevska}, {Kervella}, {Khanna}, {Kochoska}, {Kontizas}, {Kordopatis},
  {Korn}, {Kostrzewa-Rutkowska}, {Kruszy{\'n}ska}, {Lambert}, {Lanza}, {Lasne},
  {Le Campion}, {Le Fustec}, {Lebreton}, {Lebzelter}, {Leccia}, {Leclerc},
  {Lecoeur-Taibi}, {Liao}, {Licata}, {Lindstr{\o}m}, {Lister}, {Livanou},
  {Lobel}, {Madrero Pardo}, {Managau}, {Mann}, {Marchant}, {Marconi}, {Marcos
  Santos}, {Marinoni}, {Marocco}, {Marshall}, {Martin Polo},
  {Mart{\'\i}n-Fleitas}, {Masip}, {Massari}, {Mastrobuono-Battisti}, {Mazeh},
  {McMillan}, {Messina}, {Michalik}, {Millar}, {Mints}, {Molina}, {Molinaro},
  {Moln{\'a}r}, {Montegriffo}, {Mor}, {Morbidelli}, {Morel}, {Morris},
  {Mulone}, {Munoz}, {Muraveva}, {Murphy}, {Musella}, {Noval}, {Ord{\'e}novic},
  {Orr{\`u}}, {Osinde}, {Pagani}, {Pagano}, {Palaversa}, {Palicio}, {Panahi},
  {Pawlak}, {Pe{\~n}alosa Esteller}, {Penttil{\"a}}, {Piersimoni}, {Pineau},
  {Plachy}, {Plum}, {Poggio}, {Poretti}, {Poujoulet}, {Pr{\v{s}}a}, {Pulone},
  {Racero}, {Ragaini}, {Rainer}, {Raiteri}, {Rambaux}, {Ramos}, {Ramos-Lerate},
  {Re Fiorentin}, {Regibo}, {Reyl{\'e}}, {Ripepi}, {Riva}, {Rixon}, {Robichon},
  {Robin}, {Roelens}, {Rohrbasser}, {Romero-G{\'o}mez}, {Rowell}, {Royer},
  {Rybicki}, {Sadowski}, {Sagrist{\`a} Sell{\'e}s}, {Sahlmann}, {Salgado},
  {Salguero}, {Samaras}, {Sanchez Gimenez}, {Sanna}, {Santove{\~n}a},
  {Sarasso}, {Schultheis}, {Sciacca}, {Segol}, {Segovia}, {S{\'e}gransan},
  {Semeux}, {Shahaf}, {Siddiqui}, {Siebert}, {Siltala}, {Slezak}, {Smart},
  {Solano}, {Solitro}, {Souami}, {Souchay}, {Spagna}, {Spoto}, {Steele},
  {Steidelm{\"u}ller}, {Stephenson}, {S{\"u}veges}, {Szabados}, {Szegedi-Elek},
  {Taris}, {Tauran}, {Taylor}, {Teixeira}, {Thuillot}, {Tonello}, {Torra},
  {Torra}, {Turon}, {Unger}, {Vaillant}, {van Dillen}, {Vanel}, {Vecchiato},
  {Viala}, {Vicente}, {Voutsinas}, {Weiler}, {Wevers}, {Wyrzykowski}, {Yoldas},
  {Yvard}, {Zhao}, {Zorec}, {Zucker}, {Zurbach}, \&
  {Zwitter}}]{2021A&A...649A...1G}
{Gaia Collaboration}, {Brown}, A.~G.~A., {Vallenari}, A., {et~al.}
  2021{\natexlab{a}}, \aap, 649, A1

\bibitem[{{Gaia Collaboration} {et~al.}(2016{\natexlab{a}}){Gaia
  Collaboration}, {Brown}, {Vallenari}, {Prusti}, {de Bruijne}, {Mignard},
  {Drimmel}, {Babusiaux}, {Bailer-Jones}, {Bastian}, {Biermann}, {Evans},
  {Eyer}, {Jansen}, {Jordi}, {Katz}, {Klioner}, {Lammers}, {Lindegren}, {Luri},
  {O'Mullane}, {Panem}, {Pourbaix}, {Randich}, {Sartoretti}, {Siddiqui},
  {Soubiran}, {Valette}, {van Leeuwen}, {Walton}, {Aerts}, {Arenou}, {Cropper},
  {H{\o}g}, {Lattanzi}, {Grebel}, {Holland}, {Huc}, {Passot}, {Perryman},
  {Bramante}, {Cacciari}, {Casta{\~n}eda}, {Chaoul}, {Cheek}, {De Angeli},
  {Fabricius}, {Guerra}, {Hern{\'a}ndez}, {Jean-Antoine-Piccolo}, {Masana},
  {Messineo}, {Mowlavi}, {Nienartowicz}, {Ord{\'o}{\~n}ez-Blanco}, {Panuzzo},
  {Portell}, {Richards}, {Riello}, {Seabroke}, {Tanga}, {Th{\'e}venin},
  {Torra}, {Els}, {Gracia-Abril}, {Comoretto}, {Garcia-Reinaldos}, {Lock},
  {Mercier}, {Altmann}, {Andrae}, {Astraatmadja}, {Bellas-Velidis}, {Benson},
  {Berthier}, {Blomme}, {Busso}, {Carry}, {Cellino}, {Clementini}, {Cowell},
  {Creevey}, {Cuypers}, {Davidson}, {De Ridder}, {de Torres}, {Delchambre},
  {Dell'Oro}, {Ducourant}, {Fr{\'e}mat}, {Garc{\'\i}a-Torres}, {Gosset},
  {Halbwachs}, {Hambly}, {Harrison}, {Hauser}, {Hestroffer}, {Hodgkin},
  {Huckle}, {Hutton}, {Jasniewicz}, {Jordan}, {Kontizas}, {Korn}, {Lanzafame},
  {Manteiga}, {Moitinho}, {Muinonen}, {Osinde}, {Pancino}, {Pauwels}, {Petit},
  {Recio-Blanco}, {Robin}, {Sarro}, {Siopis}, {Smith}, {Smith}, {Sozzetti},
  {Thuillot}, {van Reeven}, {Viala}, {Abbas}, {Abreu Aramburu}, {Accart},
  {Aguado}, {Allan}, {Allasia}, {Altavilla}, {{\'A}lvarez}, {Alves},
  {Anderson}, {Andrei}, {Anglada Varela}, {Antiche}, {Antoja}, {Ant{\'o}n},
  {Arcay}, {Bach}, {Baker}, {Balaguer-N{\'u}{\~n}ez}, {Barache}, {Barata},
  {Barbier}, {Barblan}, {Barrado y Navascu{\'e}s}, {Barros}, {Barstow},
  {Becciani}, {Bellazzini}, {Bello Garc{\'\i}a}, {Belokurov}, {Bendjoya},
  {Berihuete}, {Bianchi}, {Bienaym{\'e}}, {Billebaud}, {Blagorodnova},
  {Blanco-Cuaresma}, {Boch}, {Bombrun}, {Borrachero}, {Bouquillon}, {Bourda},
  {Bouy}, {Bragaglia}, {Breddels}, {Brouillet}, {Br{\"u}semeister},
  {Bucciarelli}, {Burgess}, {Burgon}, {Burlacu}, {Busonero}, {Buzzi}, {Caffau},
  {Cambras}, {Campbell}, {Cancelliere}, {Cantat-Gaudin}, {Carlucci},
  {Carrasco}, {Castellani}, {Charlot}, {Charnas}, {Chiavassa}, {Clotet},
  {Cocozza}, {Collins}, {Costigan}, {Crifo}, {Cross}, {Crosta}, {Crowley},
  {Dafonte}, {Damerdji}, {Dapergolas}, {David}, {David}, {De Cat}, {de Felice},
  {de Laverny}, {De Luise}, {De March}, {de Martino}, {de Souza}, {Debosscher},
  {del Pozo}, {Delbo}, {Delgado}, {Delgado}, {Di Matteo}, {Diakite},
  {Distefano}, {Dolding}, {Dos Anjos}, {Drazinos}, {Duran}, {Dzigan},
  {Edvardsson}, {Enke}, {Evans}, {Eynard Bontemps}, {Fabre}, {Fabrizio},
  {Faigler}, {Falc{\~a}o}, {Farr{\`a}s Casas}, {Federici}, {Fedorets},
  {Fern{\'a}ndez-Hern{\'a}ndez}, {Fernique}, {Fienga}, {Figueras}, {Filippi},
  {Findeisen}, {Fonti}, {Fouesneau}, {Fraile}, {Fraser}, {Fuchs}, {Gai},
  {Galleti}, {Galluccio}, {Garabato}, {Garc{\'\i}a-Sedano}, {Garofalo},
  {Garralda}, {Gavras}, {Gerssen}, {Geyer}, {Gilmore}, {Girona}, {Giuffrida},
  {Gomes}, {Gonz{\'a}lez-Marcos}, {Gonz{\'a}lez-N{\'u}{\~n}ez},
  {Gonz{\'a}lez-Vidal}, {Granvik}, {Guerrier}, {Guillout}, {Guiraud},
  {G{\'u}rpide}, {Guti{\'e}rrez-S{\'a}nchez}, {Guy}, {Haigron},
  {Hatzidimitriou}, {Haywood}, {Heiter}, {Helmi}, {Hobbs}, {Hofmann}, {Holl},
  {Holland}, {Hunt}, {Hypki}, {Icardi}, {Irwin}, {Jevardat de Fombelle},
  {Jofr{\'e}}, {Jonker}, {Jorissen}, {Julbe}, {Karampelas}, {Kochoska},
  {Kohley}, {Kolenberg}, {Kontizas}, {Koposov}, {Kordopatis}, {Koubsky},
  {Krone-Martins}, {Kudryashova}, {Kull}, {Bachchan}, {Lacoste-Seris}, {Lanza},
  {Lavigne}, {Le Poncin-Lafitte}, {Lebreton}, {Lebzelter}, {Leccia}, {Leclerc},
  {Lecoeur-Taibi}, {Lemaitre}, {Lenhardt}, {Leroux}, {Liao}, {Licata},
  {Lindstr{\o}m}, {Lister}, {Livanou}, {Lobel}, {L{\"o}ffler}, {L{\'o}pez},
  {Lorenz}, {MacDonald}, {Magalh{\~a}es Fernandes}, {Managau}, {Mann},
  {Mantelet}, {Marchal}, {Marchant}, {Marconi}, {Marinoni}, {Marrese},
  {Marschalk{\'o}}, {Marshall}, {Mart{\'\i}n-Fleitas}, {Martino}, {Mary},
  {Matijevi{\v{c}}}, {Mazeh}, {McMillan}, {Messina}, {Michalik}, {Millar},
  {Miranda}, {Molina}, {Molinaro}, {Molinaro}, {Moln{\'a}r}, {Moniez},
  {Montegriffo}, {Mor}, {Mora}, {Morbidelli}, {Morel}, {Morgenthaler},
  {Morris}, {Mulone}, {Muraveva}, {Musella}, {Narbonne}, {Nelemans},
  {Nicastro}, {Noval}, {Ord{\'e}novic}, {Ordieres-Mer{\'e}}, {Osborne},
  {Pagani}, {Pagano}, {Pailler}, {Palacin}, {Palaversa}, {Parsons}, {Pecoraro},
  {Pedrosa}, {Pentik{\"a}inen}, {Pichon}, {Piersimoni}, {Pineau}, {Plachy},
  {Plum}, {Poujoulet}, {Pr{\v{s}}a}, {Pulone}, {Ragaini}, {Rago}, {Rambaux},
  {Ramos-Lerate}, {Ranalli}, {Rauw}, {Read}, {Regibo}, {Reyl{\'e}}, {Ribeiro},
  {Rimoldini}, {Ripepi}, {Riva}, {Rixon}, {Roelens}, {Romero-G{\'o}mez},
  {Rowell}, {Royer}, {Ruiz-Dern}, {Sadowski}, {Sagrist{\`a} Sell{\'e}s},
  {Sahlmann}, {Salgado}, {Salguero}, {Sarasso}, {Savietto}, {Schultheis},
  {Sciacca}, {Segol}, {Segovia}, {Segransan}, {Shih}, {Smareglia}, {Smart},
  {Solano}, {Solitro}, {Sordo}, {Soria Nieto}, {Souchay}, {Spagna}, {Spoto},
  {Stampa}, {Steele}, {Steidelm{\"u}ller}, {Stephenson}, {Stoev}, {Suess},
  {S{\"u}veges}, {Surdej}, {Szabados}, {Szegedi-Elek}, {Tapiador}, {Taris},
  {Tauran}, {Taylor}, {Teixeira}, {Terrett}, {Tingley}, {Trager}, {Turon},
  {Ulla}, {Utrilla}, {Valentini}, {van Elteren}, {Van Hemelryck}, {van
  Leeuwen}, {Varadi}, {Vecchiato}, {Veljanoski}, {Via}, {Vicente}, {Vogt},
  {Voss}, {Votruba}, {Voutsinas}, {Walmsley}, {Weiler}, {Weingrill}, {Wevers},
  {Wyrzykowski}, {Yoldas}, {{\v{Z}}erjal}, {Zucker}, {Zurbach}, {Zwitter},
  {Alecu}, {Allen}, {Allende Prieto}, {Amorim}, {Anglada-Escud{\'e}},
  {Arsenijevic}, {Azaz}, {Balm}, {Beck}, {Bernstein}, {Bigot}, {Bijaoui},
  {Blasco}, {Bonfigli}, {Bono}, {Boudreault}, {Bressan}, {Brown}, {Brunet},
  {Bunclark}, {Buonanno}, {Butkevich}, {Carret}, {Carrion}, {Chemin},
  {Ch{\'e}reau}, {Corcione}, {Darmigny}, {de Boer}, {de Teodoro}, {de Zeeuw},
  {Delle Luche}, {Domingues}, {Dubath}, {Fodor}, {Fr{\'e}zouls}, {Fries},
  {Fustes}, {Fyfe}, {Gallardo}, {Gallegos}, {Gardiol}, {Gebran}, {Gomboc},
  {G{\'o}mez}, {Grux}, {Gueguen}, {Heyrovsky}, {Hoar}, {Iannicola}, {Isasi
  Parache}, {Janotto}, {Joliet}, {Jonckheere}, {Keil}, {Kim}, {Klagyivik},
  {Klar}, {Knude}, {Kochukhov}, {Kolka}, {Kos}, {Kutka}, {Lainey}, {LeBouquin},
  {Liu}, {Loreggia}, {Makarov}, {Marseille}, {Martayan}, {Martinez-Rubi},
  {Massart}, {Meynadier}, {Mignot}, {Munari}, {Nguyen}, {Nordlander}, {Ocvirk},
  {O'Flaherty}, {Olias Sanz}, {Ortiz}, {Osorio}, {Oszkiewicz}, {Ouzounis},
  {Palmer}, {Park}, {Pasquato}, {Peltzer}, {Peralta}, {P{\'e}turaud},
  {Pieniluoma}, {Pigozzi}, {Poels}, {Prat}, {Prod'homme}, {Raison}, {Rebordao},
  {Risquez}, {Rocca-Volmerange}, {Rosen}, {Ruiz-Fuertes}, {Russo}, {Sembay},
  {Serraller Vizcaino}, {Short}, {Siebert}, {Silva}, {Sinachopoulos}, {Slezak},
  {Soffel}, {Sosnowska}, {Strai{\v{z}}ys}, {ter Linden}, {Terrell}, {Theil},
  {Tiede}, {Troisi}, {Tsalmantza}, {Tur}, {Vaccari}, {Vachier}, {Valles}, {Van
  Hamme}, {Veltz}, {Virtanen}, {Wallut}, {Wichmann}, {Wilkinson}, {Ziaeepour},
  \& {Zschocke}}]{2016A&A...595A...2G}
{Gaia Collaboration}, {Brown}, A.~G.~A., {Vallenari}, A., {et~al.}
  2016{\natexlab{a}}, \aap, 595, A2

\bibitem[{{Gaia Collaboration} {et~al.}(2016{\natexlab{b}}){Gaia
  Collaboration}, {Prusti}, {de Bruijne}, {Brown}, {Vallenari}, {Babusiaux},
  {Bailer-Jones}, {Bastian}, {Biermann}, {Evans}, {Eyer}, {Jansen}, {Jordi},
  {Klioner}, {Lammers}, {Lindegren}, {Luri}, {Mignard}, {Milligan}, {Panem},
  {Poinsignon}, {Pourbaix}, {Randich}, {Sarri}, {Sartoretti}, {Siddiqui},
  {Soubiran}, {Valette}, {van Leeuwen}, {Walton}, {Aerts}, {Arenou}, {Cropper},
  {Drimmel}, {H{\o}g}, {Katz}, {Lattanzi}, {O'Mullane}, {Grebel}, {Holland},
  {Huc}, {Passot}, {Bramante}, {Cacciari}, {Casta{\~n}eda}, {Chaoul}, {Cheek},
  {De Angeli}, {Fabricius}, {Guerra}, {Hern{\'a}ndez}, {Jean-Antoine-Piccolo},
  {Masana}, {Messineo}, {Mowlavi}, {Nienartowicz}, {Ord{\'o}{\~n}ez-Blanco},
  {Panuzzo}, {Portell}, {Richards}, {Riello}, {Seabroke}, {Tanga},
  {Th{\'e}venin}, {Torra}, {Els}, {Gracia-Abril}, {Comoretto},
  {Garcia-Reinaldos}, {Lock}, {Mercier}, {Altmann}, {Andrae}, {Astraatmadja},
  {Bellas-Velidis}, {Benson}, {Berthier}, {Blomme}, {Busso}, {Carry},
  {Cellino}, {Clementini}, {Cowell}, {Creevey}, {Cuypers}, {Davidson}, {De
  Ridder}, {de Torres}, {Delchambre}, {Dell'Oro}, {Ducourant}, {Fr{\'e}mat},
  {Garc{\'\i}a-Torres}, {Gosset}, {Halbwachs}, {Hambly}, {Harrison}, {Hauser},
  {Hestroffer}, {Hodgkin}, {Huckle}, {Hutton}, {Jasniewicz}, {Jordan},
  {Kontizas}, {Korn}, {Lanzafame}, {Manteiga}, {Moitinho}, {Muinonen},
  {Osinde}, {Pancino}, {Pauwels}, {Petit}, {Recio-Blanco}, {Robin}, {Sarro},
  {Siopis}, {Smith}, {Smith}, {Sozzetti}, {Thuillot}, {van Reeven}, {Viala},
  {Abbas}, {Abreu Aramburu}, {Accart}, {Aguado}, {Allan}, {Allasia},
  {Altavilla}, {{\'A}lvarez}, {Alves}, {Anderson}, {Andrei}, {Anglada Varela},
  {Antiche}, {Antoja}, {Ant{\'o}n}, {Arcay}, {Atzei}, {Ayache}, {Bach},
  {Baker}, {Balaguer-N{\'u}{\~n}ez}, {Barache}, {Barata}, {Barbier}, {Barblan},
  {Baroni}, {Barrado y Navascu{\'e}s}, {Barros}, {Barstow}, {Becciani},
  {Bellazzini}, {Bellei}, {Bello Garc{\'\i}a}, {Belokurov}, {Bendjoya},
  {Berihuete}, {Bianchi}, {Bienaym{\'e}}, {Billebaud}, {Blagorodnova},
  {Blanco-Cuaresma}, {Boch}, {Bombrun}, {Borrachero}, {Bouquillon}, {Bourda},
  {Bouy}, {Bragaglia}, {Breddels}, {Brouillet}, {Br{\"u}semeister},
  {Bucciarelli}, {Budnik}, {Burgess}, {Burgon}, {Burlacu}, {Busonero}, {Buzzi},
  {Caffau}, {Cambras}, {Campbell}, {Cancelliere}, {Cantat-Gaudin}, {Carlucci},
  {Carrasco}, {Castellani}, {Charlot}, {Charnas}, {Charvet}, {Chassat},
  {Chiavassa}, {Clotet}, {Cocozza}, {Collins}, {Collins}, {Costigan}, {Crifo},
  {Cross}, {Crosta}, {Crowley}, {Dafonte}, {Damerdji}, {Dapergolas}, {David},
  {David}, {De Cat}, {de Felice}, {de Laverny}, {De Luise}, {De March}, {de
  Martino}, {de Souza}, {Debosscher}, {del Pozo}, {Delbo}, {Delgado},
  {Delgado}, {di Marco}, {Di Matteo}, {Diakite}, {Distefano}, {Dolding}, {Dos
  Anjos}, {Drazinos}, {Dur{\'a}n}, {Dzigan}, {Ecale}, {Edvardsson}, {Enke},
  {Erdmann}, {Escolar}, {Espina}, {Evans}, {Eynard Bontemps}, {Fabre},
  {Fabrizio}, {Faigler}, {Falc{\~a}o}, {Farr{\`a}s Casas}, {Faye}, {Federici},
  {Fedorets}, {Fern{\'a}ndez-Hern{\'a}ndez}, {Fernique}, {Fienga}, {Figueras},
  {Filippi}, {Findeisen}, {Fonti}, {Fouesneau}, {Fraile}, {Fraser}, {Fuchs},
  {Furnell}, {Gai}, {Galleti}, {Galluccio}, {Garabato}, {Garc{\'\i}a-Sedano},
  {Gar{\'e}}, {Garofalo}, {Garralda}, {Gavras}, {Gerssen}, {Geyer}, {Gilmore},
  {Girona}, {Giuffrida}, {Gomes}, {Gonz{\'a}lez-Marcos},
  {Gonz{\'a}lez-N{\'u}{\~n}ez}, {Gonz{\'a}lez-Vidal}, {Granvik}, {Guerrier},
  {Guillout}, {Guiraud}, {G{\'u}rpide}, {Guti{\'e}rrez-S{\'a}nchez}, {Guy},
  {Haigron}, {Hatzidimitriou}, {Haywood}, {Heiter}, {Helmi}, {Hobbs},
  {Hofmann}, {Holl}, {Holland}, {Hunt}, {Hypki}, {Icardi}, {Irwin}, {Jevardat
  de Fombelle}, {Jofr{\'e}}, {Jonker}, {Jorissen}, {Julbe}, {Karampelas},
  {Kochoska}, {Kohley}, {Kolenberg}, {Kontizas}, {Koposov}, {Kordopatis},
  {Koubsky}, {Kowalczyk}, {Krone-Martins}, {Kudryashova}, {Kull}, {Bachchan},
  {Lacoste-Seris}, {Lanza}, {Lavigne}, {Le Poncin-Lafitte}, {Lebreton},
  {Lebzelter}, {Leccia}, {Leclerc}, {Lecoeur-Taibi}, {Lemaitre}, {Lenhardt},
  {Leroux}, {Liao}, {Licata}, {Lindstr{\o}m}, {Lister}, {Livanou}, {Lobel},
  {L{\"o}ffler}, {L{\'o}pez}, {Lopez-Lozano}, {Lorenz}, {Loureiro},
  {MacDonald}, {Magalh{\~a}es Fernandes}, {Managau}, {Mann}, {Mantelet},
  {Marchal}, {Marchant}, {Marconi}, {Marie}, {Marinoni}, {Marrese},
  {Marschalk{\'o}}, {Marshall}, {Mart{\'\i}n-Fleitas}, {Martino}, {Mary},
  {Matijevi{\v{c}}}, {Mazeh}, {McMillan}, {Messina}, {Mestre}, {Michalik},
  {Millar}, {Miranda}, {Molina}, {Molinaro}, {Molinaro}, {Moln{\'a}r},
  {Moniez}, {Montegriffo}, {Monteiro}, {Mor}, {Mora}, {Morbidelli}, {Morel},
  {Morgenthaler}, {Morley}, {Morris}, {Mulone}, {Muraveva}, {Musella},
  {Narbonne}, {Nelemans}, {Nicastro}, {Noval}, {Ord{\'e}novic},
  {Ordieres-Mer{\'e}}, {Osborne}, {Pagani}, {Pagano}, {Pailler}, {Palacin},
  {Palaversa}, {Parsons}, {Paulsen}, {Pecoraro}, {Pedrosa}, {Pentik{\"a}inen},
  {Pereira}, {Pichon}, {Piersimoni}, {Pineau}, {Plachy}, {Plum}, {Poujoulet},
  {Pr{\v{s}}a}, {Pulone}, {Ragaini}, {Rago}, {Rambaux}, {Ramos-Lerate},
  {Ranalli}, {Rauw}, {Read}, {Regibo}, {Renk}, {Reyl{\'e}}, {Ribeiro},
  {Rimoldini}, {Ripepi}, {Riva}, {Rixon}, {Roelens}, {Romero-G{\'o}mez},
  {Rowell}, {Royer}, {Rudolph}, {Ruiz-Dern}, {Sadowski}, {Sagrist{\`a}
  Sell{\'e}s}, {Sahlmann}, {Salgado}, {Salguero}, {Sarasso}, {Savietto},
  {Schnorhk}, {Schultheis}, {Sciacca}, {Segol}, {Segovia}, {Segransan},
  {Serpell}, {Shih}, {Smareglia}, {Smart}, {Smith}, {Solano}, {Solitro},
  {Sordo}, {Soria Nieto}, {Souchay}, {Spagna}, {Spoto}, {Stampa}, {Steele},
  {Steidelm{\"u}ller}, {Stephenson}, {Stoev}, {Suess}, {S{\"u}veges}, {Surdej},
  {Szabados}, {Szegedi-Elek}, {Tapiador}, {Taris}, {Tauran}, {Taylor},
  {Teixeira}, {Terrett}, {Tingley}, {Trager}, {Turon}, {Ulla}, {Utrilla},
  {Valentini}, {van Elteren}, {Van Hemelryck}, {van Leeuwen}, {Varadi},
  {Vecchiato}, {Veljanoski}, {Via}, {Vicente}, {Vogt}, {Voss}, {Votruba},
  {Voutsinas}, {Walmsley}, {Weiler}, {Weingrill}, {Werner}, {Wevers},
  {Whitehead}, {Wyrzykowski}, {Yoldas}, {{\v{Z}}erjal}, {Zucker}, {Zurbach},
  {Zwitter}, {Alecu}, {Allen}, {Allende Prieto}, {Amorim},
  {Anglada-Escud{\'e}}, {Arsenijevic}, {Azaz}, {Balm}, {Beck}, {Bernstein},
  {Bigot}, {Bijaoui}, {Blasco}, {Bonfigli}, {Bono}, {Boudreault}, {Bressan},
  {Brown}, {Brunet}, {Bunclark}, {Buonanno}, {Butkevich}, {Carret}, {Carrion},
  {Chemin}, {Ch{\'e}reau}, {Corcione}, {Darmigny}, {de Boer}, {de Teodoro}, {de
  Zeeuw}, {Delle Luche}, {Domingues}, {Dubath}, {Fodor}, {Fr{\'e}zouls},
  {Fries}, {Fustes}, {Fyfe}, {Gallardo}, {Gallegos}, {Gardiol}, {Gebran},
  {Gomboc}, {G{\'o}mez}, {Grux}, {Gueguen}, {Heyrovsky}, {Hoar}, {Iannicola},
  {Isasi Parache}, {Janotto}, {Joliet}, {Jonckheere}, {Keil}, {Kim},
  {Klagyivik}, {Klar}, {Knude}, {Kochukhov}, {Kolka}, {Kos}, {Kutka}, {Lainey},
  {LeBouquin}, {Liu}, {Loreggia}, {Makarov}, {Marseille}, {Martayan},
  {Martinez-Rubi}, {Massart}, {Meynadier}, {Mignot}, {Munari}, {Nguyen},
  {Nordlander}, {Ocvirk}, {O'Flaherty}, {Olias Sanz}, {Ortiz}, {Osorio},
  {Oszkiewicz}, {Ouzounis}, {Palmer}, {Park}, {Pasquato}, {Peltzer}, {Peralta},
  {P{\'e}turaud}, {Pieniluoma}, {Pigozzi}, {Poels}, {Prat}, {Prod'homme},
  {Raison}, {Rebordao}, {Risquez}, {Rocca-Volmerange}, {Rosen}, {Ruiz-Fuertes},
  {Russo}, {Sembay}, {Serraller Vizcaino}, {Short}, {Siebert}, {Silva},
  {Sinachopoulos}, {Slezak}, {Soffel}, {Sosnowska}, {Strai{\v{z}}ys}, {ter
  Linden}, {Terrell}, {Theil}, {Tiede}, {Troisi}, {Tsalmantza}, {Tur},
  {Vaccari}, {Vachier}, {Valles}, {Van Hamme}, {Veltz}, {Virtanen}, {Wallut},
  {Wichmann}, {Wilkinson}, {Ziaeepour}, \& {Zschocke}}]{2016A&A...595A...1G}
{Gaia Collaboration}, {Prusti}, T., {de Bruijne}, J.~H.~J., {et~al.}
  2016{\natexlab{b}}, \aap, 595, A1

\bibitem[{{Gaia Collaboration} {et~al.}(2021{\natexlab{b}}){Gaia
  Collaboration}, {Smart}, {Sarro}, {Rybizki}, {Reyl{\'e}}, {Robin}, {Hambly},
  {Abbas}, {Barstow}, {de Bruijne}, {Bucciarelli}, {Carrasco}, {Cooper},
  {Hodgkin}, {Masana}, {Michalik}, {Sahlmann}, {Sozzetti}, {Brown},
  {Vallenari}, {Prusti}, {Babusiaux}, {Biermann}, {Creevey}, {Evans}, {Eyer},
  {Hutton}, {Jansen}, {Jordi}, {Klioner}, {Lammers}, {Lindegren}, {Luri},
  {Mignard}, {Panem}, {Pourbaix}, {Randich}, {Sartoretti}, {Soubiran},
  {Walton}, {Arenou}, {Bailer-Jones}, {Bastian}, {Cropper}, {Drimmel}, {Katz},
  {Lattanzi}, {van Leeuwen}, {Bakker}, {Casta{\~n}eda}, {De Angeli},
  {Ducourant}, {Fabricius}, {Fouesneau}, {Fr{\'e}mat}, {Guerra}, {Guerrier},
  {Guiraud}, {Jean-Antoine Piccolo}, {Messineo}, {Mowlavi}, {Nicolas},
  {Nienartowicz}, {Pailler}, {Panuzzo}, {Riclet}, {Roux}, {Seabroke}, {Sordo},
  {Tanga}, {Th{\'e}venin}, {Gracia-Abril}, {Portell}, {Teyssier}, {Altmann},
  {Andrae}, {Bellas-Velidis}, {Benson}, {Berthier}, {Blomme}, {Brugaletta},
  {Burgess}, {Busso}, {Carry}, {Cellino}, {Cheek}, {Clementini}, {Damerdji},
  {Davidson}, {Delchambre}, {Dell'Oro}, {Fern{\'a}ndez-Hern{\'a}ndez},
  {Galluccio}, {Garc{\'\i}a-Lario}, {Garcia-Reinaldos},
  {Gonz{\'a}lez-N{\'u}{\~n}ez}, {Gosset}, {Haigron}, {Halbwachs}, {Harrison},
  {Hatzidimitriou}, {Heiter}, {Hern{\'a}ndez}, {Hestroffer}, {Holl},
  {Jan{\ss}en}, {Jevardat de Fombelle}, {Jordan}, {Krone-Martins}, {Lanzafame},
  {L{\"o}ffler}, {Lorca}, {Manteiga}, {Marchal}, {Marrese}, {Moitinho}, {Mora},
  {Muinonen}, {Osborne}, {Pancino}, {Pauwels}, {Recio-Blanco}, {Richards},
  {Riello}, {Rimoldini}, {Roegiers}, {Siopis}, {Smith}, {Ulla}, {Utrilla}, {van
  Leeuwen}, {van Reeven}, {Abreu Aramburu}, {Accart}, {Aerts}, {Aguado},
  {Ajaj}, {Altavilla}, {{\'A}lvarez}, {{\'A}lvarez Cid-Fuentes}, {Alves},
  {Anderson}, {Anglada Varela}, {Antoja}, {Audard}, {Baines}, {Baker},
  {Balaguer-N{\'u}{\~n}ez}, {Balbinot}, {Balog}, {Barache}, {Barbato},
  {Barros}, {Bartolom{\'e}}, {Bassilana}, {Bauchet}, {Baudesson-Stella},
  {Becciani}, {Bellazzini}, {Bernet}, {Bertone}, {Bianchi}, {Blanco-Cuaresma},
  {Boch}, {Bombrun}, {Bossini}, {Bouquillon}, {Bragaglia}, {Bramante},
  {Breedt}, {Bressan}, {Brouillet}, {Burlacu}, {Busonero}, {Butkevich},
  {Buzzi}, {Caffau}, {Cancelliere}, {C{\'a}novas}, {Cantat-Gaudin}, {Carballo},
  {Carlucci}, {Carnerero}, {Casamiquela}, {Castellani}, {Castro-Ginard},
  {Castro Sampol}, {Chaoul}, {Charlot}, {Chemin}, {Chiavassa}, {Cioni},
  {Comoretto}, {Cornez}, {Cowell}, {Crifo}, {Crosta}, {Crowley}, {Dafonte},
  {Dapergolas}, {David}, {David}, {de Laverny}, {De Luise}, {De March}, {De
  Ridder}, {de Souza}, {de Teodoro}, {de Torres}, {del Peloso}, {del Pozo},
  {Delgado}, {Delgado}, {Delisle}, {Di Matteo}, {Diakite}, {Diener},
  {Distefano}, {Dolding}, {Eappachen}, {Edvardsson}, {Enke}, {Esquej}, {Fabre},
  {Fabrizio}, {Faigler}, {Fedorets}, {Fernique}, {Fienga}, {Figueras},
  {Fouron}, {Fragkoudi}, {Fraile}, {Franke}, {Gai}, {Garabato},
  {Garcia-Gutierrez}, {Garc{\'\i}a-Torres}, {Garofalo}, {Gavras}, {Gerlach},
  {Geyer}, {Giacobbe}, {Gilmore}, {Girona}, {Giuffrida}, {Gomel}, {Gomez},
  {Gonzalez-Santamaria}, {Gonz{\'a}lez-Vidal}, {Granvik},
  {Guti{\'e}rrez-S{\'a}nchez}, {Guy}, {Hauser}, {Haywood}, {Helmi}, {Hidalgo},
  {Hilger}, {H{\l}adczuk}, {Hobbs}, {Holland}, {Huckle}, {Jasniewicz},
  {Jonker}, {Juaristi Campillo}, {Julbe}, {Karbevska}, {Kervella}, {Khanna},
  {Kochoska}, {Kontizas}, {Kordopatis}, {Korn}, {Kostrzewa-Rutkowska},
  {Kruszy{\'n}ska}, {Lambert}, {Lanza}, {Lasne}, {Le Campion}, {Le Fustec},
  {Lebreton}, {Lebzelter}, {Leccia}, {Leclerc}, {Lecoeur-Taibi}, {Liao},
  {Licata}, {Lindstr{\o}m}, {Lister}, {Livanou}, {Lobel}, {Madrero Pardo},
  {Managau}, {Mann}, {Marchant}, {Marconi}, {Marcos Santos}, {Marinoni},
  {Marocco}, {Marshall}, {Martin Polo}, {Mart{\'\i}n-Fleitas}, {Masip},
  {Massari}, {Mastrobuono-Battisti}, {Mazeh}, {McMillan}, {Messina}, {Millar},
  {Mints}, {Molina}, {Molinaro}, {Moln{\'a}r}, {Montegriffo}, {Mor},
  {Morbidelli}, {Morel}, {Morris}, {Mulone}, {Munoz}, {Muraveva}, {Murphy},
  {Musella}, {Noval}, {Ord{\'e}novic}, {Orr{\`u}}, {Osinde}, {Pagani},
  {Pagano}, {Palaversa}, {Palicio}, {Panahi}, {Pawlak}, {Pe{\~n}alosa
  Esteller}, {Penttil{\"a}}, {Piersimoni}, {Pineau}, {Plachy}, {Plum},
  {Poggio}, {Poretti}, {Poujoulet}, {Pr{\v{s}}a}, {Pulone}, {Racero},
  {Ragaini}, {Rainer}, {Raiteri}, {Rambaux}, {Ramos}, {Ramos-Lerate}, {Re
  Fiorentin}, {Regibo}, {Ripepi}, {Riva}, {Rixon}, {Robichon}, {Robin},
  {Roelens}, {Rohrbasser}, {Romero-G{\'o}mez}, {Rowell}, {Royer}, {Rybicki},
  {Sadowski}, {Sagrist{\`a} Sell{\'e}s}, {Salgado}, {Salguero}, {Samaras},
  {Sanchez Gimenez}, {Sanna}, {Santove{\~n}a}, {Sarasso}, {Schultheis},
  {Sciacca}, {Segol}, {Segovia}, {S{\'e}gransan}, {Semeux}, {Shahaf},
  {Siddiqui}, {Siebert}, {Siltala}, {Slezak}, {Solano}, {Solitro}, {Souami},
  {Souchay}, {Spagna}, {Spoto}, {Steele}, {Steidelm{\"u}ller}, {Stephenson},
  {S{\"u}veges}, {Szabados}, {Szegedi-Elek}, {Taris}, {Tauran}, {Taylor},
  {Teixeira}, {Thuillot}, {Tonello}, {Torra}, {Torra}, {Turon}, {Unger},
  {Vaillant}, {van Dillen}, {Vanel}, {Vecchiato}, {Viala}, {Vicente},
  {Voutsinas}, {Weiler}, {Wevers}, {Wyrzykowski}, {Yoldas}, {Yvard}, {Zhao},
  {Zorec}, {Zucker}, {Zurbach}, \& {Zwitter}}]{2021A&A...649A...6G}
{Gaia Collaboration}, {Smart}, R.~L., {Sarro}, L.~M., {et~al.}
  2021{\natexlab{b}}, \aap, 649, A6

\bibitem[{{Hart}(1979)}]{1979Icar...37..351H}
{Hart}, M.~H. 1979, \icarus, 37, 351

\bibitem[{{Hartman} \& {L{\'e}pine}(2020)}]{2020ApJS..247...66H}
{Hartman}, Z.~D. \& {L{\'e}pine}, S. 2020, \apjs, 247, 66

\bibitem[{{Haywood}(2001)}]{2001MNRAS.325.1365H}
{Haywood}, M. 2001, \mnras, 325, 1365

\bibitem[{{Houdebine} {et~al.}(2019){Houdebine}, {Mullan}, {Doyle}, {de La
  Vieuville}, {Butler}, \& {Paletou}}]{2019AJ....158...56H}
{Houdebine}, {\'E}.~R., {Mullan}, D.~J., {Doyle}, J.~G., {et~al.} 2019, \aj,
  158, 56

\bibitem[{{Ivanov} {et~al.}(2013){Ivanov}, {Minniti}, {Hempel}, {Kurtev},
  {Toledo}, {Saito}, {Alonso-Garc{\'\i}a}, {Beam{\'\i}n}, {Borissova},
  {Catelan}, {Chen{\'e}}, {Emerson}, {Gonz{\'a}lez}, {Lucas}, {Mart{\'\i}n},
  {Rejkuba}, \& {Gromadzki}}]{2013A&A...560A..21I}
{Ivanov}, V.~D., {Minniti}, D., {Hempel}, M., {et~al.} 2013, \aap, 560, A21

\bibitem[{{Kasting} {et~al.}(1993){Kasting}, {Whitmire}, \&
  {Reynolds}}]{1993Icar..101..108K}
{Kasting}, J.~F., {Whitmire}, D.~P., \& {Reynolds}, R.~T. 1993, \icarus, 101,
  108

\bibitem[{{Kesseli} {et~al.}(2017){Kesseli}, {West}, {Veyette}, {Harrison},
  {Feldman}, \& {Bochanski}}]{2017ApJS..230...16K}
{Kesseli}, A.~Y., {West}, A.~A., {Veyette}, M., {et~al.} 2017, \apjs, 230, 16

\bibitem[{{Kirkpatrick} {et~al.}(2010){Kirkpatrick}, {Looper}, {Burgasser},
  {Schurr}, {Cutri}, {Cushing}, {Cruz}, {Sweet}, {Knapp}, {Barman},
  {Bochanski}, {Roellig}, {McLean}, {McGovern}, \&
  {Rice}}]{2010ApJS..190..100K}
{Kirkpatrick}, J.~D., {Looper}, D.~L., {Burgasser}, A.~J., {et~al.} 2010,
  \apjs, 190, 100

\bibitem[{{Kopparapu}(2013)}]{2013ApJ...767L...8K}
{Kopparapu}, R.~K. 2013, \apjl, 767, L8

\bibitem[{{Kowalski} {et~al.}(2009){Kowalski}, {Hawley}, {Hilton}, {Becker},
  {West}, {Bochanski}, \& {Sesar}}]{2009AJ....138..633K}
{Kowalski}, A.~F., {Hawley}, S.~L., {Hilton}, E.~J., {et~al.} 2009, \aj, 138,
  633

\bibitem[{{Kuznetsov} {et~al.}(2019){Kuznetsov}, {del Burgo}, {Pavlenko}, \&
  {Frith}}]{2019ApJ...878..134K}
{Kuznetsov}, M.~K., {del Burgo}, C., {Pavlenko}, Y.~V., \& {Frith}, J. 2019,
  \apj, 878, 134

\bibitem[{{L{\'e}pine} \& {Gaidos}(2011)}]{2011AJ....142..138L}
{L{\'e}pine}, S. \& {Gaidos}, E. 2011, \aj, 142, 138

\bibitem[{{Luyten}(1980)}]{1980nltt.bookQ....L}
{Luyten}, W.~J. 1980, {NLTT Catalogue. Volume\_III. 0\_\_to -30\_.}

\bibitem[{{Mann} {et~al.}(2019){Mann}, {Dupuy}, {Kraus}, {Gaidos}, {Ansdell},
  {Ireland}, {Rizzuto}, {Hung}, {Dittmann}, {Factor}, {Feiden}, {Martinez},
  {Ru{\'\i}z-Rodr{\'\i}guez}, \& {Thao}}]{2019ApJ...871...63M}
{Mann}, A.~W., {Dupuy}, T., {Kraus}, A.~L., {et~al.} 2019, \apj, 871, 63

\bibitem[{{Minniti}(2018)}]{2018ASSP...51...63M}
{Minniti}, D. 2018, in The Vatican Observatory, Castel Gandolfo: 80th
  Anniversary Celebration, ed. G.~{Gionti} \& J.-B. {Kikwaya Eluo}, Vol.~51, 63

\bibitem[{{Mowlavi} {et~al.}(2021){Mowlavi}, {Rimoldini}, {Evans}, {Riello},
  {De Angeli}, {Palaversa}, {Audard}, {Eyer}, {Garcia-Lario}, {Gavras}, {Holl},
  {Jevardat de Fombelle}, {Lec{\oe}ur-Ta{\"\i}bi}, \&
  {Nienartowicz}}]{2021A&A...648A..44M}
{Mowlavi}, N., {Rimoldini}, L., {Evans}, D.~W., {et~al.} 2021, \aap, 648, A44

\bibitem[{{Mugrauer} \& {Michel}(2020)}]{2020AN....341..996M}
{Mugrauer}, M. \& {Michel}, K.-U. 2020, Astronomische Nachrichten, 341, 996

\bibitem[{{Muirhead} {et~al.}(2011){Muirhead}, {Edelstein}, {Erskine},
  {Wright}, {Muterspaugh}, {Covey}, {Wishnow}, {Hamren}, {Andelson}, {Kimber},
  {Mercer}, {Halverson}, {Vanderburg}, {Mondo}, {Czeszumska}, \&
  {Lloyd}}]{2011PASP..123..709M}
{Muirhead}, P.~S., {Edelstein}, J., {Erskine}, D.~J., {et~al.} 2011, \pasp,
  123, 709

\bibitem[{{Ochsenbein} {et~al.}(2000){Ochsenbein}, {Bauer}, \&
  {Marcout}}]{2000A&AS..143...23O}
{Ochsenbein}, F., {Bauer}, P., \& {Marcout}, J. 2000, \aaps, 143, 23

\bibitem[{{Pecaut} \& {Mamajek}(2013)}]{2013ApJS..208....9P}
{Pecaut}, M.~J. \& {Mamajek}, E.~E. 2013, \apjs, 208, 9

\bibitem[{{Phan-Bao} {et~al.}(2008){Phan-Bao}, {Bessell}, {Mart{\'\i}n},
  {Simon}, {Borsenberger}, {Tata}, {Guibert}, {Crifo}, {Forveille}, {Delfosse},
  {Lim}, \& {de Batz}}]{2008MNRAS.383..831P}
{Phan-Bao}, N., {Bessell}, M.~S., {Mart{\'\i}n}, E.~L., {et~al.} 2008, \mnras,
  383, 831

\bibitem[{{Reid} {et~al.}(2004){Reid}, {Cruz}, {Allen}, {Mungall}, {Kilkenny},
  {Liebert}, {Hawley}, {Fraser}, {Covey}, {Lowrance}, {Kirkpatrick}, \&
  {Burgasser}}]{2004AJ....128..463R}
{Reid}, I.~N., {Cruz}, K.~L., {Allen}, P., {et~al.} 2004, \aj, 128, 463

\bibitem[{{Reid} {et~al.}(2001){Reid}, {van Wyk}, {Marang}, {Roberts},
  {Kilkenny}, \& {Mahoney}}]{2001MNRAS.325..931R}
{Reid}, I.~N., {van Wyk}, F., {Marang}, F., {et~al.} 2001, \mnras, 325, 931

\bibitem[{{Reyl{\'e}}(2018)}]{2018A&A...619L...8R}
{Reyl{\'e}}, C. 2018, \aap, 619, L8

\bibitem[{{Reyl{\'e}} {et~al.}(2021){Reyl{\'e}}, {Jardine}, {Fouqu{\'e}},
  {Caballero}, {Smart}, \& {Sozzetti}}]{2021A&A...650A.201R}
{Reyl{\'e}}, C., {Jardine}, K., {Fouqu{\'e}}, P., {et~al.} 2021, \aap, 650,
  A201

\bibitem[{{Rojas-Ayala} {et~al.}(2014){Rojas-Ayala}, {Iglesias}, {Minniti},
  {Saito}, \& {Surot}}]{2014A&A...571A..36R}
{Rojas-Ayala}, B., {Iglesias}, D., {Minniti}, D., {Saito}, R.~K., \& {Surot},
  F. 2014, \aap, 571, A36

\bibitem[{{Schlafly} {et~al.}(2018){Schlafly}, {Green}, {Lang}, {Daylan},
  {Finkbeiner}, {Lee}, {Meisner}, {Schlegel}, \&
  {Valdes}}]{2018ApJS..234...39S}
{Schlafly}, E.~F., {Green}, G.~M., {Lang}, D., {et~al.} 2018, \apjs, 234, 39

\bibitem[{{Smart} {et~al.}(2017){Smart}, {Marocco}, {Caballero}, {Jones},
  {Barrado}, {Beam{\'\i}n}, {Pinfield}, \& {Sarro}}]{2017MNRAS.469..401S}
{Smart}, R.~L., {Marocco}, F., {Caballero}, J.~A., {et~al.} 2017, \mnras, 469,
  401

\bibitem[{{Smart} {et~al.}(2019){Smart}, {Marocco}, {Sarro}, {Barrado},
  {Beam{\'\i}n}, {Caballero}, \& {Jones}}]{2019MNRAS.485.4423S}
{Smart}, R.~L., {Marocco}, F., {Sarro}, L.~M., {et~al.} 2019, \mnras, 485, 4423

\bibitem[{{Smith} {et~al.}(2015){Smith}, {Lucas}, {Contreras Pe{\~n}a},
  {Kurtev}, {Marocco}, {Jones}, {Beamin}, {Napiwotzki}, {Borissova},
  {Burningham}, {Faherty}, {Pinfield}, {Gromadzki}, {Ivanov}, {Minniti},
  {Stimson}, \& {Villanueva}}]{2015MNRAS.454.4476S}
{Smith}, L.~C., {Lucas}, P.~W., {Contreras Pe{\~n}a}, C., {et~al.} 2015,
  \mnras, 454, 4476

\bibitem[{{Smith} {et~al.}(2018){Smith}, {Lucas}, {Kurtev}, {Smart}, {Minniti},
  {Borissova}, {Jones}, {Zhang}, {Marocco}, {Contreras Pe{\~n}a}, {Gromadzki},
  {Kuhn}, {Drew}, {Pinfield}, \& {Bedin}}]{2018MNRAS.474.1826S}
{Smith}, L.~C., {Lucas}, P.~W., {Kurtev}, R., {et~al.} 2018, \mnras, 474, 1826

\bibitem[{{Taylor}(2005)}]{2005ASPC..347...29T}
{Taylor}, M.~B. 2005, in Astronomical Society of the Pacific Conference Series,
  Vol. 347, Astronomical Data Analysis Software and Systems XIV, ed.
  P.~{Shopbell}, M.~{Britton}, \& R.~{Ebert}, 29

\bibitem[{{Thompson} {et~al.}(2013){Thompson}, {Kirkpatrick}, {Mace},
  {Cushing}, {Gelino}, {Griffith}, {Skrutskie}, {Eisenhardt}, {Wright},
  {Marsh}, {Mix}, {Beichman}, {Faherty}, {Toloza}, {Ferrara}, {Apodaca},
  {McLean}, \& {Bloom}}]{2013PASP..125..809T}
{Thompson}, M.~A., {Kirkpatrick}, J.~D., {Mace}, G.~N., {et~al.} 2013, \pasp,
  125, 809

\bibitem[{{Tian} {et~al.}(2020){Tian}, {El-Badry}, {Rix}, \&
  {Gould}}]{2020ApJS..246....4T}
{Tian}, H.-J., {El-Badry}, K., {Rix}, H.-W., \& {Gould}, A. 2020, \apjs, 246, 4

\bibitem[{{Vacca} {et~al.}(2003){Vacca}, {Cushing}, \&
  {Rayner}}]{2003PASP..115..389V}
{Vacca}, W.~D., {Cushing}, M.~C., \& {Rayner}, J.~T. 2003, \pasp, 115, 389

\bibitem[{{Wenger} {et~al.}(2000){Wenger}, {Ochsenbein}, {Egret}, {Dubois},
  {Bonnarel}, {Borde}, {Genova}, {Jasniewicz}, {Lalo{\"e}}, {Lesteven}, \&
  {Monier}}]{2000A&AS..143....9W}
{Wenger}, M., {Ochsenbein}, F., {Egret}, D., {et~al.} 2000, \aaps, 143, 9

\bibitem[{{West} {et~al.}(2011){West}, {Morgan}, {Bochanski}, {Andersen},
  {Bell}, {Kowalski}, {Davenport}, {Hawley}, {Schmidt}, {Bernat}, {Hilton},
  {Muirhead}, {Covey}, {Rojas-Ayala}, {Schlawin}, {Gooding}, {Schluns},
  {Dhital}, {Pineda}, \& {Jones}}]{2011AJ....141...97W}
{West}, A.~A., {Morgan}, D.~P., {Bochanski}, J.~J., {et~al.} 2011, \aj, 141, 97

\bibitem[{{Winters} {et~al.}(2021){Winters}, {Charbonneau}, {Henry}, {Irwin},
  {Jao}, {Riedel}, \& {Slatten}}]{2021AJ....161...63W}
{Winters}, J.~G., {Charbonneau}, D., {Henry}, T.~J., {et~al.} 2021, \aj, 161,
  63

\bibitem[{{Wright} {et~al.}(2010){Wright}, {Eisenhardt}, {Mainzer}, {Ressler},
  {Cutri}, {Jarrett}, {Kirkpatrick}, {Padgett}, {McMillan}, {Skrutskie},
  {Stanford}, {Cohen}, {Walker}, {Mather}, {Leisawitz}, {Gautier}, {McLean},
  {Benford}, {Lonsdale}, {Blain}, {Mendez}, {Irace}, {Duval}, {Liu}, {Royer},
  {Heinrichsen}, {Howard}, {Shannon}, {Kendall}, {Walsh}, {Larsen}, {Cardon},
  {Schick}, {Schwalm}, {Abid}, {Fabinsky}, {Naes}, \&
  {Tsai}}]{2010AJ....140.1868W}
{Wright}, E.~L., {Eisenhardt}, P. R.~M., {Mainzer}, A.~K., {et~al.} 2010, \aj,
  140, 1868

\bibitem[{{Yershov}(2014)}]{2014Ap&SS.354...97Y}
{Yershov}, V.~N. 2014, \apss, 354, 97

\end{thebibliography}

%\Online

\begin{appendix} %First online appendix
\section{Spectral energy distributions of the 99 low-mass objects}

Spectral energy distributions with the model fit of each object included in this work are available online at \url{https://github.com/andreamejiasb/SEDs_VVVX-GCNS_stars}

\end{appendix}

\end{document}